\documentclass[iop]{emulateapj-rtx4}
\usepackage{amssymb}
\usepackage{amsmath}
\usepackage{comment}
\usepackage{hyperref}
\providecommand{\cm[1]}{\ensuremath{\,{\rm cm}^{#1}}}
\providecommand{\kms}{\ensuremath{\,{\rm km\,s}^{-1}}}
\providecommand{\Ang}{\ensuremath{\,\mbox{\AA}}}
\providecommand{\kpc}{\ensuremath{\,\mathrm{kpc}}}
\providecommand{\Lya}{\ensuremath{\mathrm{Ly}\alpha}} 
\providecommand{\Cthr}{\ensuremath{\mathrm{C\,IV}}}

\providecommand{\zciv}{\ensuremath{z_{1548}}}
\providecommand{\zqso}{\ensuremath{z_{\rm QSO}}}
\providecommand{\dv[1]}{\ensuremath{\delta v_{#1}}}
\providecommand{\dvqso}{\ensuremath{\dv{\rm QSO}}}
\providecommand{\dvem}{\ensuremath{\dv{\rm em}}}
\providecommand{\dvabs}{\ensuremath{\dv{\rm abs}}}
\providecommand{\dvciv}{\ensuremath{\dv{\rm C\,IV}}}
\providecommand{\EWr}{\ensuremath{W_{\rm r}}}
\providecommand{\sigEWr}{\ensuremath{\sigma_{\EWr}}}
\providecommand{\EWlin[1]}{\ensuremath{\EWr{}_{,\mathrm{#1}}}}

\providecommand{\N[1]}{\ensuremath{N_{\rm #1}}}
\providecommand{\NCIV}{\ensuremath{\N{\Cthr}}}
\providecommand{\sigNCIV}{\ensuremath{\sigma_{\N{\Cthr}}}}
\providecommand{\logCIV}{\ensuremath{\log \NCIV}}

\providecommand{\OmCIV}{\ensuremath{\Omega_{\Cthr}}}
\providecommand{\sigOmCIV}{\ensuremath{\sigma_{\Omega}}}
\providecommand{\ff[1]}{\ensuremath{f(#1)}}

\providecommand{\Num}{\ensuremath{\mathcal{N}}}

\providecommand{\DXp}{\ensuremath{\Delta X}}
\providecommand{\sigDX}{\ensuremath{\sigma_{\DXp}}}
\providecommand{\DX[1]}{\ensuremath{\DXp(#1)}}
\providecommand{\ud}{\ensuremath{d}} 
\providecommand{\dXdz}{\ensuremath{\ud X/\ud z}}
\providecommand{\dNCIVdz}{\ensuremath{\ud \Num_{\mathrm{C\,IV}}/\ud z}}
\providecommand{\dNCIVdX}{\ensuremath{\ud \Num_{\mathrm{C\,IV}}/\ud X}}
\providecommand{\dNFkIVdX}{\ensuremath{\ud \Num_{\mathrm{afp}}/\ud X}}
\providecommand{\dNMgIIdX}{\ensuremath{\ud \Num_{\mathrm{Mg\,II}}/\ud
  X}}
\providecommand{\sigphys}{\ensuremath{\sigma_{\rm phys}}}
\providecommand{\ncom}{\ensuremath{n_{\rm com}}}
\providecommand{\ncomuv}{\ensuremath{n_{\rm com,UV}}}

\begin{document}


\title{Precious Metals in SDSS Quasar Spectra I:
  Tracking the Evolution of \\ Strong, $1.5 < z < 4.5$ \ion{C}{4}
  Absorbers with Thousands of Systems}

\author{Kathy L. Cooksey\altaffilmark{1,6}, Melodie
  M. Kao\altaffilmark{2}, Robert A. Simcoe\altaffilmark{3}, John M.
  O'Meara\altaffilmark{4}, and J. Xavier Prochaska\altaffilmark{5}}

\altaffiltext{1}{MIT Kavli Institute for Astrophysics \& Space
  Research, 77 Massachusetts 
    Avenue, 37-685, Cambridge, MA 02139, USA; kcooksey@space.mit.edu}
\altaffiltext{2}{Caltech, MC 249-17,  1200 East California Boulevard,
  Pasadena, CA 91125; mkao@caltech.edu}
\altaffiltext{3}{Department of Physics, MIT, 77 Massachusetts
    Avenue, 37-664D, Cambridge, MA 02139, USA;
  simcoe@space.mit.edu}  
\altaffiltext{4}{Department of Chemistry and Physics, Saint Michael's
  College, One Winooski Park, Colchester, VT 05439; jomeara@smcvt.edu}  
\altaffiltext{5}{Department of Astronomy \& UCO/Lick Observatory,
 University of California, 1156 High Street, Santa Cruz, CA 95064, USA;
 xavier@ucolick.org}
\altaffiltext{6}{NSF Astronomy \& Astrophysics Postdoctoral Fellow}

\shorttitle{Evolution of Strong \ion{C}{4} Absorbers}\shortauthors{Cooksey et al.}

\slugcomment{Draft 4: \today}

\begin{abstract}
  We have vastly increased the \ion{C}{4} statistics at intermediate
  redshift by surveying the thousands of quasars in the Sloan Digital Sky
  Survey Data-Release 7. We visually verified over 16,000 \ion{C}{4}
  systems with $1.46 < z < 4.55$---a sample size that renders Poisson
  error negligible.  Detailed Monte Carlo simulations show we are
  approximately 50\% complete down to rest equivalent widths $\EWr \approx
  0.6\Ang$. We analyzed the sample as a whole and in ten small
  redshift bins with approximately 1500 doublets each. The equivalent
  width frequency distributions \ff{\EWr}\ were well modeled by an
  exponential, with little evolution in shape. In contrast with
  previous studies that modeled the frequency distribution as a
  single power law, the fitted exponential gives a finite mass density
  for the \ion{C}{4} ions. The co-moving line density \dNCIVdX\
  evolved smoothly with redshift, increasing by a factor of
  $2.37\pm0.09$ from $z = 4.55$ to 1.96, then plateauing at $\dNCIVdX
  \approx 0.34$ for $z = 1.96$ to 1.46. Comparing our SDSS sample with
  $z < 1$ (ultraviolet) and $z > 5$ (infrared) surveys, we see an
  approximately 10-fold increase in \dNCIVdX\ over $z \approx 6
  \rightarrow 0$, for $\EWr \ge 0.6\Ang$. This suggests a monotonic
  and significant increase in the enrichment of gas outside galaxies
  over the 12\,Gyr lifetime of the universe. 
\end{abstract}



\section{Introduction}\label{sec.intro}

\addtocounter{footnote}{6}

The study of the large-scale structure of the universe provides
top-level constraints on models of galaxy evolution. Heavy elements
are produced in the stars of galaxies. A variety of feedback
processes move these metals from the sites of production into the
intergalactic medium (IGM), enriching the material for future
generations of stars. The cosmic enrichment cycle generically refers
to the movement of gas from inside galaxies to the IGM and, possibly,
back again (perhaps many times).  The amount of heavy elements, the
number of ionizing photons from galaxies and quasars, and the spatial
distribution of material are driven by hierarchical structure
formation and galactic processes such as star formation and
feedback. Spectroscopic surveys of quasars yield a random sample of
intervening absorbing gas clouds that can be used to constrain the
on-going and summative enrichment processes in the universe.

\ion{C}{4} $\lambda\lambda1548,1550$ doublets are important tracers of
the IGM and its evolution from $z \approx 6$ to today
\citep{steidel90, barlowandtytler98, ellisonetal00, songaila01,
  boksenbergetal03ph, schayeetal03, scannapiecoetal06,
  danforthandshull08, ryanweberetal09, beckeretal09, cookseyetal10,
  dodoricoetal10, simcoeetal11}. This transition has been
well-studied at $1.5 \lesssim z \lesssim 5.5$ for the following
reasons. First, it is a strong transition of a common metal. Second,
it is observable outside the Ly$\alpha$ forest, where it becomes
easier to identify. Third, it redshifts into optical passbands at $z
= 1.5$. Lastly, it is a resonant doublet, which gives it
distinctive characteristics that enable surveys to be largely
automated.

Observations of \ion{C}{4} doublets constrain the cumulative effect of
the cosmic enrichment cycle. More specifically, the number and
strength of \ion{C}{4} absorbers are affected by: the amount of carbon
produced by all previous generations of stars; the spatial
distribution of the element, driven by feedback processes; and the
total ionizing radiation available to maintain the triply-ionized
transition.

The intermediate-redshift \ion{C}{4} studies have traditionally found
that the doublets follow a power law in the column density
distribution function, with a slope of $\alpha \approx -1.8$,
throughout the redshift range \citep{songaila01, boksenbergetal03ph,
  scannapiecoetal06, dodoricoetal10}. The earlier studies also
measured a roughly constant \ion{C}{4} mass density for $2 < z < 4.5$
and $12 \lesssim \logCIV \lesssim 15$ absorbers \citep{songaila01,
  boksenbergetal03ph, songaila05}. Improved observations, pushing to
bluer wavelengths (and hence lower redshift), have shown that \OmCIV,
the \ion{C}{4} mass density relative to the critical density, actually
increases smoothly from $z = 4 \rightarrow 1.5$ \citep{dodoricoetal10}
and maps well onto the, $z < 1$ values, measured with {\it Hubble
  Space Telescope} ({\it HST}) ultraviolet spectra
\citep{cookseyetal10}. Thus, \OmCIV\ increases by, approximately, a
factor of four over $z \approx 3 \rightarrow 0$ while \dNCIVdX, the
co-moving \ion{C}{4} line density, increases by a factor of two,  roughly.

Early infrared spectroscopy, which probes the $z \gtrsim 5$ universe,
first resulted in a continuation of the roughly constant \OmCIV\ out
to $z \approx 6$ \citep{ryanweberetal06, simcoe06}. However, with
limited sightlines, these studies were highly susceptible to cosmic
variance, as shown by later IR surveys, which reported that \OmCIV\
actually dropped at $z \approx 6$ \citep{ryanweberetal09,
  beckeretal09}. The latest and largest IR survey showed that \OmCIV\
drops by, approximately, a factor of four over $z \approx 4 \rightarrow 6$
\citep{simcoeetal11}, and high-redshift quasars are continuing to be
observed with the new Folded-port InfraRed Echellete (FIRE) on the
Magellan\slash Baade Telescope \citep{simcoeetal10}. In addition,
\citet{simcoeetal11} found that all previous high-redshift
measurements overestimated \OmCIV\ by $\approx30\%$, since their
lower-resolution IR spectra led to the lower-redshift \ion{Mg}{2}
$\lambda\lambda2796,2803$ being mis-identified as a strong,
high-redshift \ion{C}{4} system.

The recent low (UV) and high (IR) redshift publications
\citep{cookseyetal10, simcoeetal11} have led us to assess the state of
the intermediate (optical) redshift field \citep[e.g.,][]{songaila01,
  boksenbergetal03ph, scannapiecoetal06, dodoricoetal10}. The various
studies have disparities in their definition of an absorber,
completeness corrections, sensitivity limits, and\slash or adopted
cosmology.  Ideally, there should be a large, uniformly constructed,
$0 < z < 6$ sample in order to evaluate the evolution of the
\ion{C}{4} absorbers. We aim to produce an intermediate redshift
catalog that is fairly comparable to the recent $z < 1$ and $z > 5$
catalogs.

For our $1.5 \lesssim z \lesssim 4.5$ survey, we use more than 26,000
quasar spectra in the Sloan Digital Sky Survey
\citep[SDSS;][]{yorketal00} Data-Release 7 (DR7) database
\citep{abazajianetal09, schneideretal10}, thus making this study the
largest \ion{C}{4} survey---both in path length and in number of
absorbers---to date. Others have mined SDSS quasar spectra for
\ion{H}{1} \citep[e.g.,][]{prochaskaetal05, pierietal10lya},
\ion{Mg}{2} \citep[e.g.,][]{nestoretal05, prochteretal06}, \ion{Ca}{2}
$\lambda\lambda3934,3969$ \citep[e.g.,][]{wildetal06, cherinkaetal11},
and \ion{O}{6} $\lambda,\lambda1031,1037$ \citep[e.g.,][]{franketal10,
  pierietal10ovi}. However, this is the first time that anyone has
systematically searched for \ion{C}{4}, likely because it is more
difficult and labor intensive, due to the increased amount of
blending. This is the first in a series of papers on metals in SDSS
quasar spectra, where we will assemble and analyze self-consistent
catalogs of \ion{Si}{4} $\lambda\lambda1393,1402$, \ion{Mg}{2}, and
\ion{Ca}{2} absorbers. These species are commonly studied in quasar
absorption-line spectroscopy and typically trace the gas closest to
galaxies.

We explain how we construct our \ion{C}{4} sample in
\S\ref{sec.sample} and our completeness corrections in
\S\ref{sec.complete}. The main results are detailed in
\S\ref{sec.results}, and the discussion and summary are in
\S\S\ref{sec.discuss}--\ref{sec.summ}.  We adopt the WMAP5 cosmology:
$H_{0}=71.9\kms\,{\rm Mpc}$, $\Omega_{\rm M} = 0.258$, and
$\Omega_{\Lambda} = 0.742$ \citep{komatsuetal09}.


\begin{deluxetable*}{ll}
\tablewidth{0pt}
\tabletypesize{\scriptsize}
\tablecaption{Sightline Selection\label{tab.obssum}  }
\tablehead{
\colhead{Number} & \colhead{Description} 
}
\startdata
105783 & SDSS DR7 QSO catalog \citep{schneideretal10} \\
99569 & Excluding 6214 objects in BAL QSO catalog \citep{shenetal11} \\
48260 & Covering $\max[1310\Ang(1+\zqso),\ 3820\Ang] \le  \lambda <
\min[1548\Ang(1+\zqso)(1+\dvqso/c),\ 9200\Ang]$\tablenotemark{a} \\
26168 & With $\zqso \ge 1.7$ and $\langle {\rm S/N} \rangle \ge
4\,{\rm pixel}^{-1}$ in above wavelength range\tablenotemark{b} \\
26030 & Excluding 138 visual BAL QSOs \\
10861 & With confirmed \ion{C}{4} doublets 
\enddata
\tablenotetext{a}{The 1310\Ang\ limit excludes the contaminating
  region in the O\,I, Si\,II ``forest.'' The fiducial value
  for $\dvqso = -3000\kms$.}
\tablenotetext{b}{The details of these specific sightlines are given
  in Table \ref{tab.los}.}
\end{deluxetable*}

\begin{figure}[hbt]
  \begin{center}
  \includegraphics[width=0.47\textwidth]{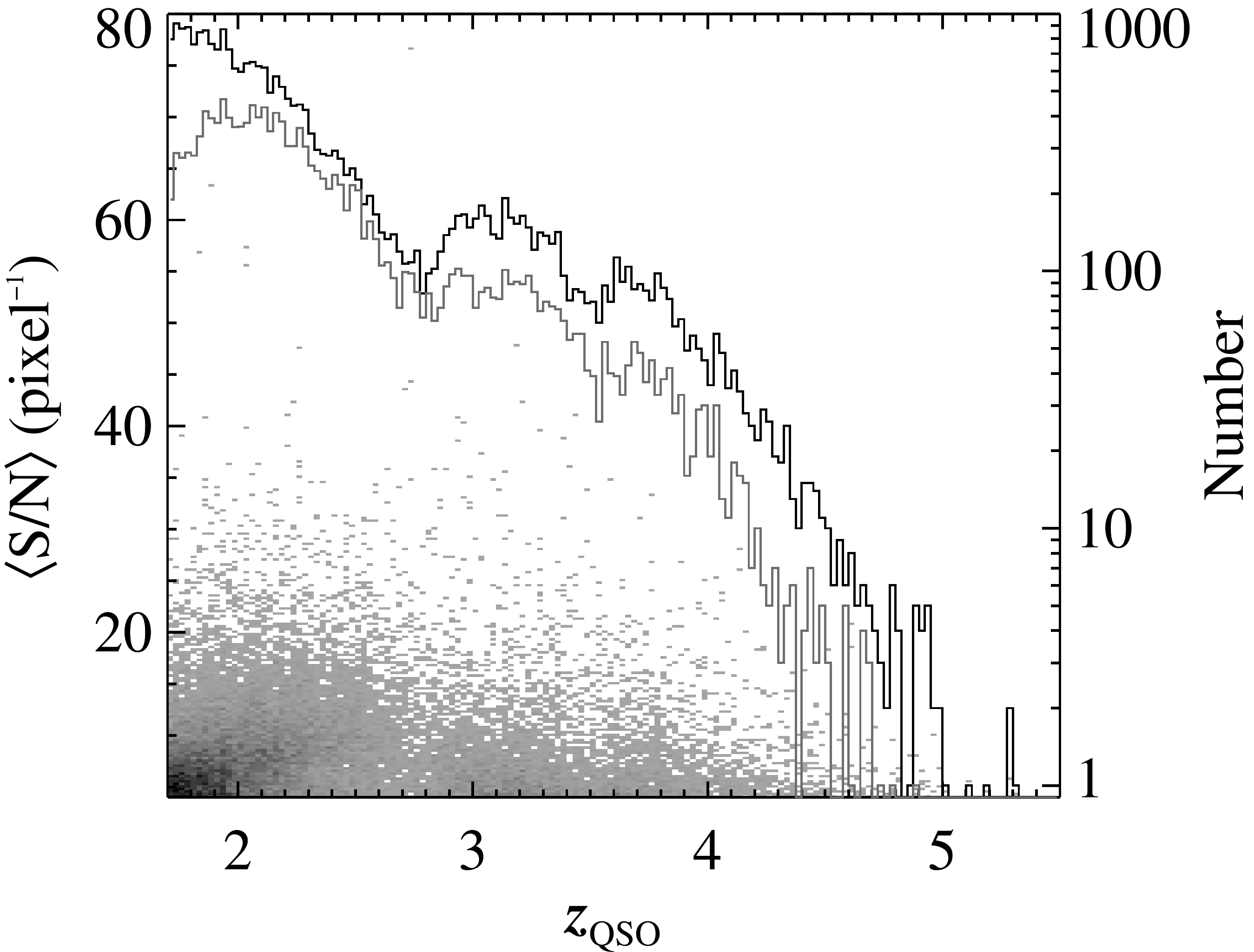} 
  \end{center}
  \caption[Redshift and $\langle {\rm S/N} \rangle$ distribution of sightlines]
  {Redshift and $\langle {\rm S/N} \rangle$ distribution of
    sightlines. Using the left-hand axis, the 2D histogram shows the
    median signal-to-noise and \zqso\ space of the analyzed 26,030 spectra. The
    black and gray histograms give the redshift distribution for all
    spectra and for the 10,861 with confirmed \ion{C}{4} absorbers,
    respectively (right-hand axis).
    \label{fig.zqsosnr}
  }
\end{figure}

\begin{deluxetable*}{llllrccccc}
\tablewidth{0pc}
\tablecaption{Sightline Summary\label{tab.los}}
\tabletypesize{\scriptsize}
\tablehead{ 
\colhead{(1)} & \colhead{(2)} & \colhead{(3)} & 
\colhead{(4)} & \colhead{(5)} & 
\colhead{(6)} & \colhead{(7)} & \colhead{(8)} & 
\colhead{(9)} & \colhead{(10)} \\ 
\colhead{QSO ID} & \colhead{R.A.} & 
\colhead{Decl.} & \colhead{\zqso} & 
\colhead{$\langle {\rm S/N} \rangle$} & \colhead{$f_{\rm BAL}$} & 
\colhead{$\Delta X_{\rm max}$} & 
\colhead{$\Num_{\rm cand}$} & \colhead{$\Num_{\rm C\,IV}$} & 
\colhead{$\delta X_{\rm C,IV}$} \\ 
 & & & & 
\colhead{(pixel$^{-1}$)} &  &  &  & & 
} 
\startdata 
52235-0750-082 & 00:00:09.38 &  +13:56:18.4 &  2.2342 &  5.94 &   0 &   1.33 &   0 &   0 & \nodata \\ 
52143-0650-199 & 00:00:09.42 & --10:27:51.9 &  1.8449 &  4.68 &   0 &   0.83 &   1 &   1 & 0.028 \\ 
52203-0685-198 & 00:00:14.82 & --01:10:30.7 &  1.8877 &  5.16 &   0 &   1.24 &   0 &   0 & \nodata \\ 
52203-0685-439 & 00:00:15.47 &  +00:52:46.8 &  1.8516 &  8.64 &   0 &   2.16 &   0 &   0 & \nodata \\ 
54389-2822-315 & 00:00:24.83 &  +24:57:03.3 &  3.2137 &  9.21 &   0 &   1.48 &   0 &   0 & \nodata \\ 
51791-0387-167 & 00:00:39.00 & --00:18:03.9 &  2.1249 &  6.61 &   0 &   1.75 &   0 &   0 & \nodata \\ 
52143-0650-178 & 00:00:50.60 & --10:21:55.9 &  2.6404 &  8.68 &   0 &   1.56 &   4 &   2 & 0.083 \\ 
52203-0685-154 & 00:00:53.17 & --00:17:32.9 &  2.7571 &  7.55 &   0 &   1.15 &   0 &   0 & \nodata \\ 
52902-1091-546 & 00:00:57.58 &  +01:06:58.6 &  2.5551 & 11.83 &   0 &   1.78 &   3 &   2 & 0.084 \\ 
51791-0387-093 & 00:00:58.22 & --00:46:46.5 &  1.8973 & 13.27 &   0 &   0.84 &   1 &   1 & 0.037 \\ 
52203-0685-134 & 00:01:25.14 &  +00:00:09.4 &  1.9739 &  5.32 &   0 &   1.52 &   0 &   0 & \nodata
\enddata 
\tablecomments{ 
Column 1 is the adopted QSO identifier from the spectroscopic modified Julian date, plate, and fiber number.
Columns 2 through 4 are from the DR7 QSO catalog \citep{schneideretal10}.
Column 5 is the median S/N measured in the region searched for C\,IV.
The binary BAL flag $f_{\rm BAL}$ in Column 6 indicates which sightlines were considered BALs by at least one author (4) and which were confirmed by the authors as BALs to exclude (8).
Column 7 is the maximum co-moving pathlength available in the sightline.
Columns 8 and 9 give the number of candidate and confirmed C\,IV doublets, respectively.
Column 10 is the pathlength blocked by the $\Num_{\rm C\,IV}$ doublets in the sightline.
(This table is available in a machine-readable form in the online journal.
A portion is shown here for guidance regarding its form and content.)
} 
\end{deluxetable*} 

\section{Constructing the \ion{C}{4} Sample}\label{sec.sample}

\subsection{SDSS Data-Release 7 Quasars}\label{subsec.dr7}

We began our survey with the 105,783 sightlines in the SDSS DR7 QSO
catalog \citet[][also see \citealt{richardsetal02} for the SDSS
quasar selection]{schneideretal10}. SDSS spectra have wavelength
coverage $3820\Ang\ \le \lambda \le 9200\Ang$ and resolution varying
from $R = 1850$ to 2200 (or 162\kms\ to 136\kms), and the reduced
spectra are binned to a log-linear scale of $69\kms\,{\rm
  pixel}^{-1}$. We immediately excluded the 6,214 broad-absorption line
(BAL) QSOs detailed in \citet{shenetal11}. 

We limited the sightlines to those with $\zqso \ge 1.7$ and median
signal-to-noise $\langle {\rm S/N} \rangle \ge 4\,{\rm pixel}^{-1}$ in
the wavelength range sensitive to \ion{C}{4} absorbers (see Figure
\ref{fig.zqsosnr}).  This range depended on the quasar redshift and the
SDSS wavelength coverage. 

Previous, smaller surveys for \ion{C}{4} doublets have searched from
the \Lya\ $\lambda1215$ to \ion{C}{4} emission wavelengths. However,
we removed a portion of this path length close to the \Lya\ emission to
avoid possible confusion between the \ion{C}{4} doublet and any
\ion{O}{1} $\lambda1302$, \ion{Si}{2} $\lambda1304$ pair.  These
latter two have a wavelength separation similar enough to the
\ion{C}{4} doublet that automated search algorithms naturally bring in
a large fraction of false positives. Since we had so much available
path length available, we simply excluded a comfortable region around
the \ion{O}{1}, \ion{Si}{2} ``forest'' in addition to the \Lya\ forest
(i.e., $\lambda_{\rm r} \ge 1310\Ang$ in the restframe of the
quasar). To exclude absorbers intrinsic to the QSO or affected by
local ionization, clustering, and\slash or enrichment, we set the
upper wavelength bound to
$1548\Ang(1+\zqso)(1+\dvqso/c)$,\footnote{Velocity offsets are defined
  as $\delta v = c(z - z_{\rm ref})/(1 + z_{\rm ref})$.} where we
initially used $\dvqso = -3000\kms$, but for the main analyses, the
limit is $\dvqso = -5000\kms$.

After the redshift, $\langle {\rm S/N} \rangle$, and coverage cuts, we
had 26,168 sightlines to search (see Table \ref{tab.obssum}). Later, we
excluded 138 sightlines as ``visual BAL'' QSOs for a final count of
26,030 spectra to analyze.

\begin{figure*}[hbt]
  \begin{center}$
    \begin{array}{cc}
  \includegraphics[width=0.47\textwidth]{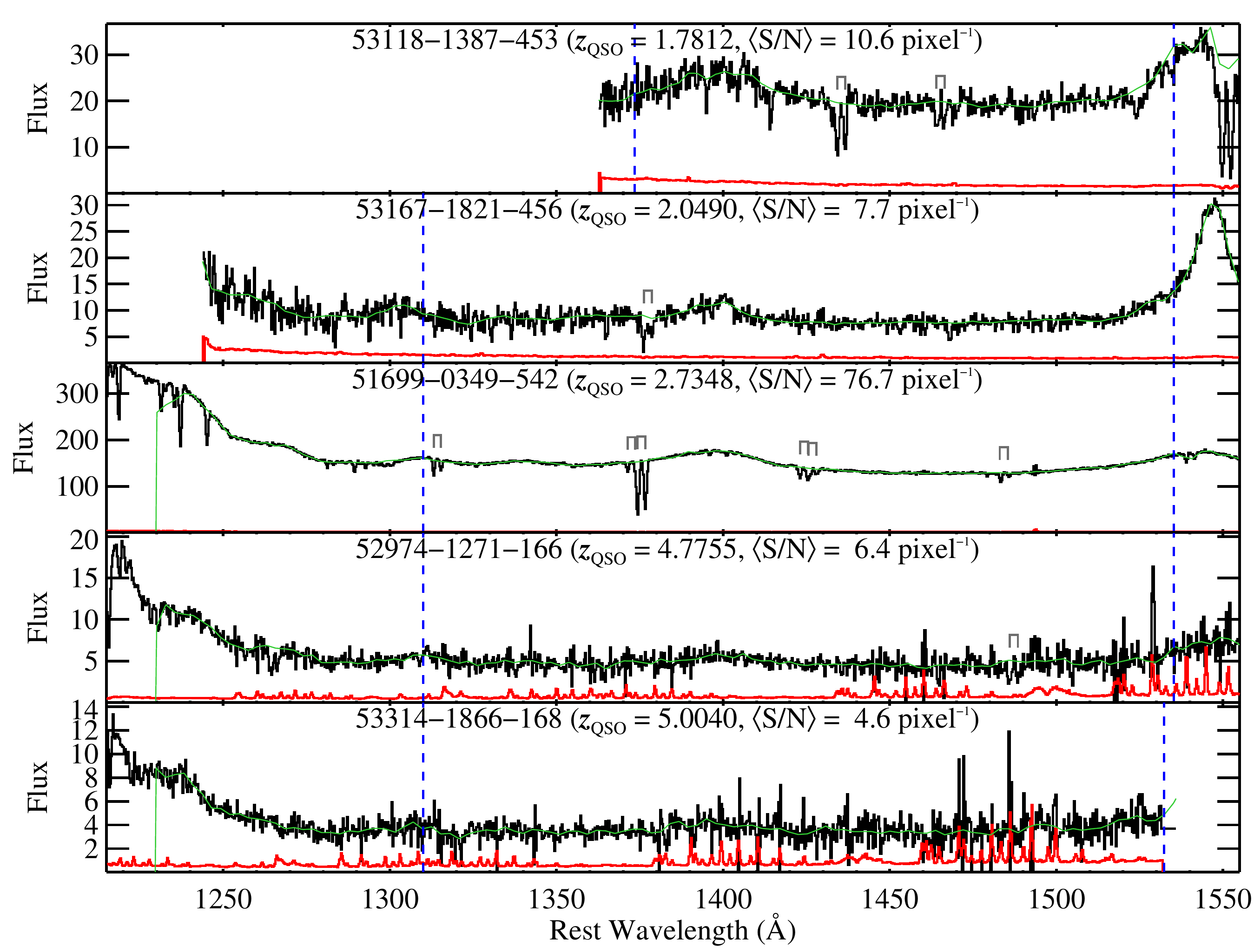}  & 
  \includegraphics[width=0.47\textwidth]{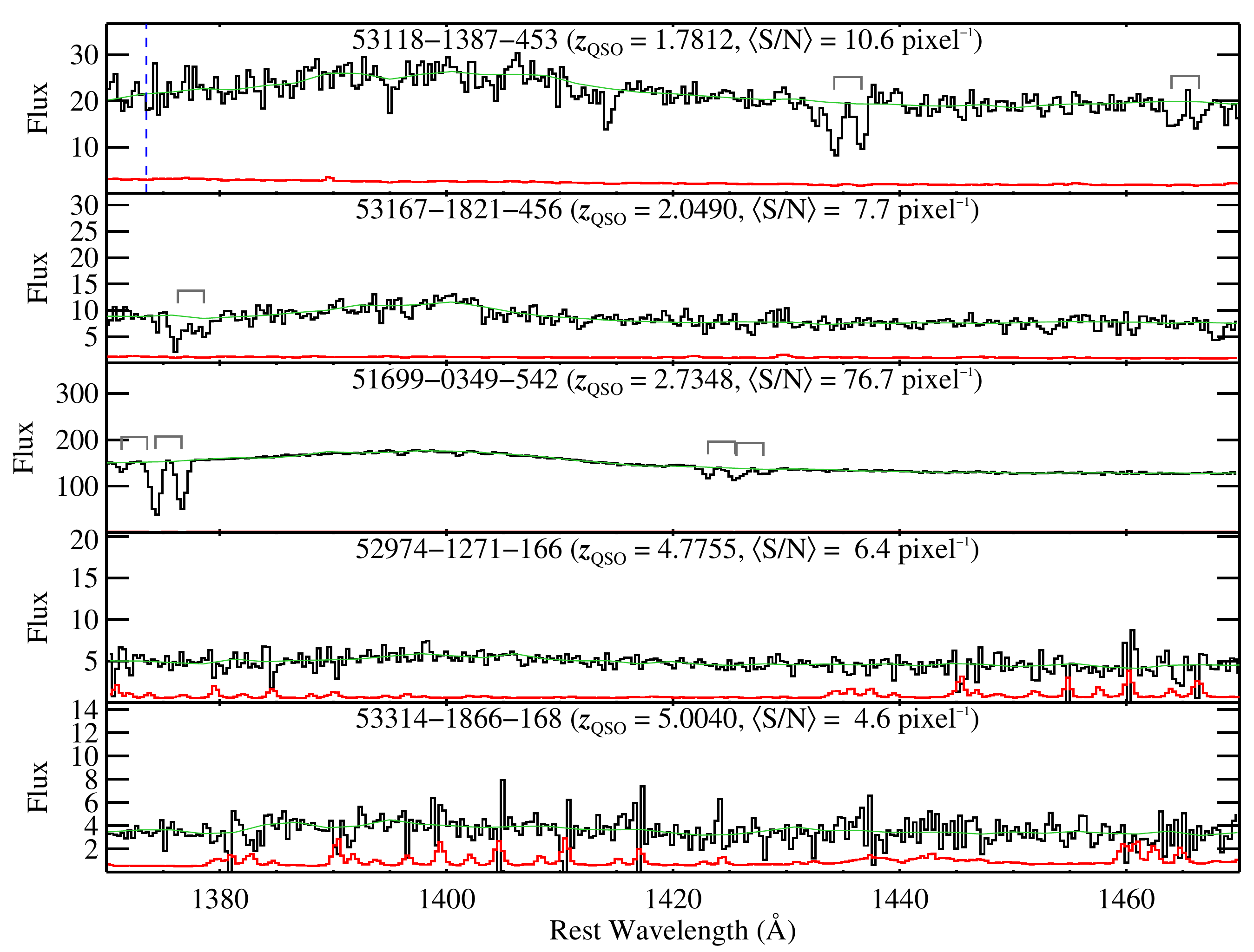} 
  \end{array}
 $\end{center}
  \caption[Examples of automated continuum fits.]
  {Examples of automated continuum fits. Several QSO spectra (black)
    and 1-$\sigma$ error (red) are shown with their automated
    continuum fits (green). On the left are the full spectra, and a
    zoom of the central wavelength range is given on the right. All
    spectra were fit for $\lambda_{\rm r} \ge 1230\Ang$ with SDSS
    eigenspectra. For $\langle {\rm S/N} \rangle \ge 8\,{\rm
      pixel}^{-1}$ spectra, the eigen-fit was adjusted with a b-spline fit
    to the normalized spectrum. We surveyed the spectra for \ion{C}{4}
    absorbers between the (blue) vertical lines, which are outside the
    \ion{O}{1}, \ion{Si}{2} ``forest'' and more than 3000\kms\
    blueward of the QSO. The gray brackets indicate real \ion{C}{4}
    doublets from the final sample.
    \label{fig.conti}
  }
\end{figure*}

\subsection{Automated Continuum Fitting}\label{subsec.conti}

We generated individual quasar continua for all $\approx26,000$
spectra in the sample using an algorithm which combined
principle-component analysis (PCA) fits, low-frequency b-spline
correction, and automated outlier pixel and absorption-line
exclusion. Since we are interested in \ion{C}{4} lines only, we
limited the fit to $\lambda_{\rm r} \ge 1230\Ang$, redward of the
\Lya\ forest. We discuss details of the method below so interested
users may replicate it; the codes are also publicly available in the
{\sc xidl} software library.\footnote{See
  \url{http://ucolick.org/~xavier/IDL/}.}  Figure \ref{fig.conti}
shows example fits for eight spectra of varying $\langle {\rm S/N}
\rangle$ level.

First, every spectrum was iteratively fit using a 50-vector basis set
of PCA ``eigenspectra'' from the SDSS DR1 quasar sample
\citep{yipetal04}. The fit coefficients and their errors were
determined using the algorithms of \citet[][Equations
4--6]{connollyandszalay99}, with spectral pixels weighted by their
inverse variance when quantifying the goodness-of-fit. The formal
per-pixel error of the PCA continuum fit was negligible compared to
the spectral shot noise.

A pixel exclusion mask was then generated upon each successive fit
iteration.  The initial mask consisted of the lowest 30\% of pixels in
successive 50-pixel windows, to filter strong absorption features and
avoid biasing the continuum low on the critical initial fit.  Then,
for each subsequent iteration, a fresh mask was constructed which
excluded pixels falling either less than $-2\sigma$ or more than
$3\sigma$ from the working continuum.  This mask was further modified
to include the neighboring pixels of each deviant pixel below the
working continuum and exclude regions with three consecutive pixels
above.  The latter correspond to systematically errant regions of the
fit that should actually be included, so they were unmasked
automatically.  This cycle was repeated with successive PCA fits
until: (a) the mask converged so successive fits did not change; (b)
the unmasked flux and continuum fit had median difference less than
0.001\%; or (c) ten iterations passed.

For high-S/N spectra, low-frequency residuals in the PCA-normalized
spectra become noticeable and affect our ability to search for weak
\ion{C}{4} lines in the best data.  We therefore performed a secondary
correction to spectra with $\langle {\rm S/N} \rangle \ge 8\,\mathrm{
  pixel}^{ -1 }$, re-fitting the PCA-normalized spectrum with an
additional third-order b-spline, with breakpoint spacing of 25\,pixels
(or 1725\kms).  As before, we used inverse-variance weighting and
clipped outliers using the same thresholds.  We refer below to this
PCA plus b-spline fit as the ``hybrid-continuum.''

The b-spline systematic error was estimated as the median difference
between the b-spline and the PCA-normalized (masked) spectrum in bins
matching the breakpoint spacing.  This error was typically of order
1\% of the shot noise in the data; it exceeded the formal PCA fit
error, and we added the two in quadrature to produce a total continuum
error.

The sigma clipping methods described above are effective at
identifying narrow features, instrumental artifacts, or defects in the
spectra.  However, they are not always optimal for finding true
absorption lines, which are always below the continuum; they can also
be kinematically complex and bias the fit. Since we have prior
information about the characteristic widths of cosmological
absorption, we develop methods in \S\ref{subsec.cand} to
machine-identify candidate absorption lines with these characteristics
for the actual survey.  We ran each PCA-normalized spectrum through a
single pass of the absorption-line finder, using the resulting
catalogs to generate a master absorption mask for final
hybrid-continuum fitting.

The full procedure for automated feature finding is described in
\S\ref{subsec.cand}; briefly, we convolve both the data and error
arrays with the instrumental response profile, and search for
absorption features with signal-to-noise ratio $\ge3.5$ per resolution
element.  For $\langle {\rm S/N} \rangle \le 16\,{\rm pixel}^{-1}$
spectra, we masked out all such features; in high-S/N cases, we also
enforced that the unconvolved spectral pixels at the line center must
deviate from the continuum by $>5\sigma$.  Pixels falling within
$\pm600\kms$ of such features were added to the exclusion set.  The
mask consisting of automatically identified absorption lines was fed
into the hybrid-continuum fit as a static mask.

This procedure produced excellent continuum fits for the vast majority
of spectra.  When it failed, the most common reasons were: poorly
measured \zqso, strong intrinsic absorption, broad-absorption lines,
and foreground emission-line galaxy spectra superimposed on the quasar
spectra.  Instead of fixing these cases interactively, which would compromise
our objective methodology and continuum error estimates, we chose to
leave them in the sample and let the effects be accounted in our
automated completeness and contamination tests (see \S\ref{sec.complete}).

\subsection{Automated Candidate Selection}\label{subsec.cand}

We limited our candidate search to \ion{C}{4} doublets where two
automatically detected lines were within $\pm150\kms$ of the
characteristic velocity separation $\dvciv = 498\kms$. The large
uncertainty cut allowed heavily blended lines into our candidate list,
but it also made the \ion{Mg}{2} doublet our largest contaminant since
$\dv{\rm Mg\,II} = 767\kms$.

Our automated feature-finding algorithm is based on
\citet{prochteretal06}. The hybrid-normalized flux and error arrays
were convolved with a Gaussian kernel with full-width at half-maximum
equivalent to a resolution element of the SDSS spectrograph, i.e.,
$\sigma_{\rm G} = 1\,{\rm pixel}$. The error array included the
continuum fit error. The regions where the convolved signal-to-noise
$({\rm S/N)_{conv}}$ is greater than or equal to $3.5$ per resolution element
were identified as absorption features and saved to the sightline's
line list. We masked out the few pixels ($\approx 5\Ang$) around the
strong skylines at 5579\Ang\ and 6302\Ang.

Candidate \ion{C}{4} doublets were compiled by pairing automatically
detected lines with the characteristic velocity separation $\dvciv
\pm 150\kms$ and by identifying isolated, automatically detected lines
that were broad enough to be a candidate doublet by themselves.  As
discussed in \S\ref{subsec.dr7}, the search region is 
set so that the \ion{C}{4} candidate is redward of the \ion{O}{1},
\ion{Si}{2} ``forest'' ($\lambda_{\rm r} \ge 1310\Ang$) and blueward
of the QSO by $3000\kms$.  
For any un-paired line, we re-ran the automated feature-finder with a
$2.5\sigma_{\rm conv}$ cut-off and searched for a partner candidate
1550 line.

To identify isolated, broad lines that should be candidate \ion{C}{4}
doublets, we used the automated procedure for measuring wavelength
bounds (described in \S\ref{subsec.meas}) to find lines with
widths $\dv{\rm lin} \ge 1.5\,\dvciv$.  

Ultimately, we identified 29,789 candidates, 6,346 of which were from
the broad-line search. 

\subsection{Measuring Absorber Properties}\label{subsec.meas}

The redshift, equivalent width, and column density of a line depend on
the definition of its wavelength bounds. The bounds were automatically
defined by where the convolved signal-to-noise array stopped
decreasing from the perspective of the automatically detected
centroid, and the bounds were not allowed to exceed the midpoint
between themselves and neighboring lines. Thus, for \ion{C}{4} lines,
the inner wavelength bounds were not allowed to exceed the midpoint of
the doublet.

The redshift, equivalent width, and column density errors included the
estimated error due to the hybrid-continuum. Redshifts were 
measured from the flux-weighted centroids. Equivalent widths were
measured by simply summing the absorbed flux within the bounds.

We used the apparent optical depth method
\citep[AODM;][]{savageandsembach91} to estimate column densities.
Systems that are saturated (discussed below) formally have 
column densities that are lower limits (binary flag $f_{N} =
2$). Column density measurements that are less than $3\sigma_{N}$ are
flagged as upper limits ($f_{N} = 4$). 

With doublets, we have two measurements of the column density, and we
use both to set \NCIV, as done in \citet{cookseyetal10}. For
doublets where both lines are measurements (not limits), \NCIV\ is
the inverse-variance weighted mean of the two values. If one line is a
measurement, \NCIV\ is set to its value. The column density {\it
  limit} is set to the inverse-variance weighted mean when both
doublet limits are the same kind. When the doublet limits bracket a
range, \NCIV\ is set to the un-weighted average with an error that
reflects the range ($f_{N} = 8$).  In the remaining cases, \NCIV\ is
set to the better measurement\slash limit and flagged $f_{N} = 16$.

The AODM systematically overestimates the true column densities in
low-S/N spectra \citep{foxetal05}. On the other hand, most doublets
that can be detected in the low-resolution SDSS spectra are saturated,
as evidenced by the doublet ratios of near unity; therefore, the AOD
column densities would formally be lower limits. We define the doublet
to be unsaturated when the equivalent width ratio
$\EWlin{1548}/\EWlin{1550} > 2-\sigma_{\rm R}$, where $\sigma_{\rm R}$
is the ratio error due to the uncertainties in the measured equivalent
widths. All other doublets are flagged as saturated with column
densities that are lower limits; by this criterion, approximately 84\%
of the final catalog are saturated (see Table \ref{tab.civ}).

\begin{figure}[hbt]
  \begin{center}
  \includegraphics[width=0.47\textwidth]{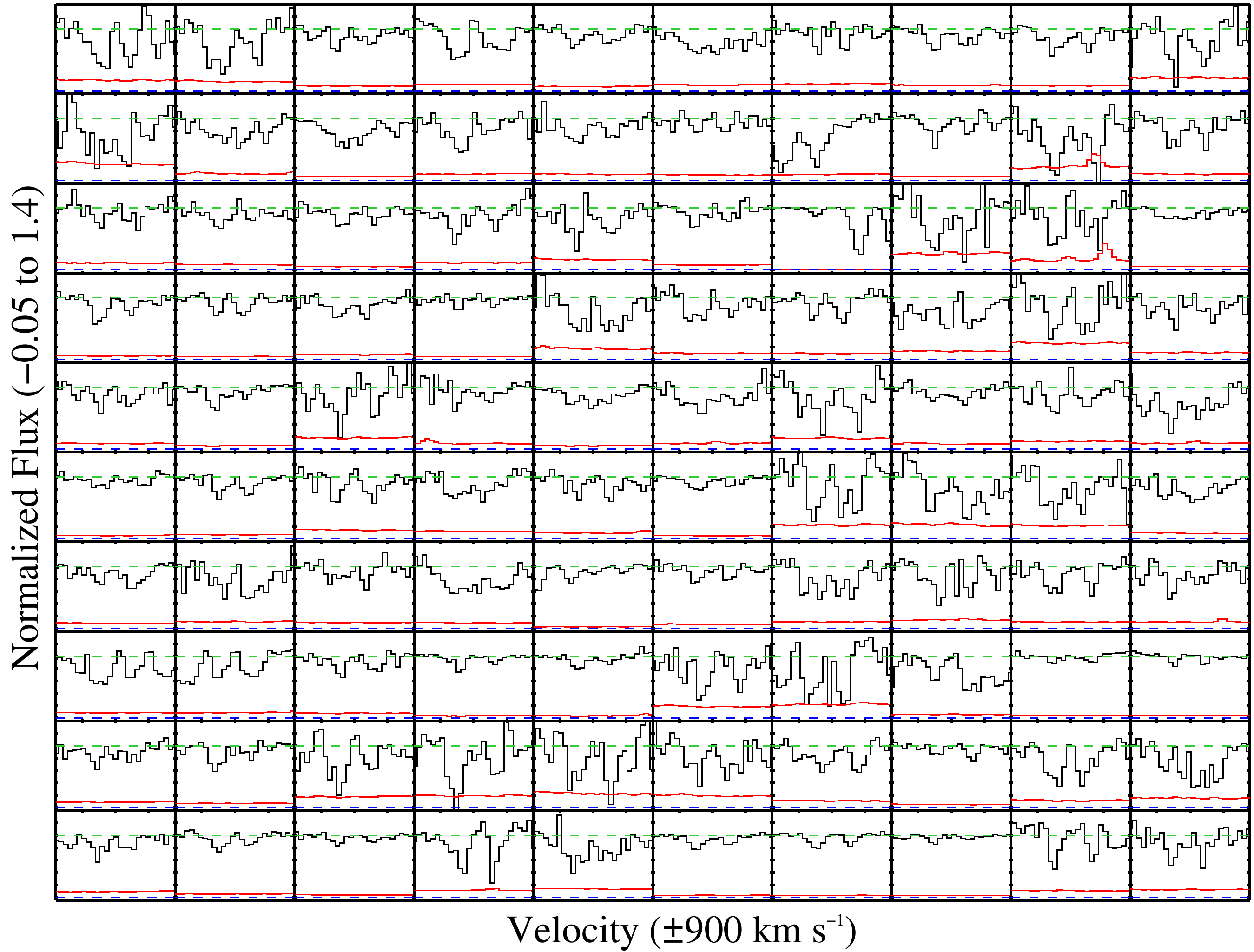} 
  \end{center}
  \caption[Example of \ion{C}{4} absorber sample.]
  {Example of \ion{C}{4} absorbers. One hundred doublet profiles are
    plotted from $-900\kms$ to 900\kms with respect to the center of
    the doublet and from $-0.05$ to 1.4 in normalized flux units. The
    flux is show in black and the 1-$\sigma$ error in red. The
    horizontal dashed lines indicate the flux at unity (green) and
    zero (blue).
    \label{fig.civexmpl}
  }
\end{figure}

\subsection{Interactive Doublet Verification}\label{subsec.verify}
 
The candidate \ion{C}{4} selection relied solely on the characteristic
wavelength separation of the doublet. However, true, unblended
\ion{C}{4} doublets have a well-defined doublet ratio
$\EWlin{1548}/\EWlin{1550}$ in the range of one to two in the
saturated and unsaturated regimes, respectively. The resolved
profiles of the doublet lines are very similar in the absence of
blending. Most \ion{C}{4} absorbers are associated with \Lya\
absorption, though ``naked'' systems may exist
\citep{schayeetal07}. There are frequently other metal lines
associated with the \ion{C}{4} absorption, such as the \ion{Si}{4}
and\slash or \ion{Mg}{2} doublets.  In addition,
outside of the \Lya\ (and \ion{O}{1},\ion{Si}{2}) forest, line
confusion and blending are less severe, and the \ion{C}{4} doublet is
one of the more common metal lines. 

We developed a graphical user interface to present all this information for each system
to assist us in rating the 29,789 candidates. We leveraged
our experience with \ion{C}{4} absorption systems and the varying
amounts of information available, depending on redshift, to assign
each candidate a rating:
\vspace{-0.5\baselineskip}
\begin{itemize}\addtolength{\itemsep}{-0.5\baselineskip}
\item[0.] definitely false---either can definitively identify
  the lines as other metal lines or as spurious bad pixels
  masquerading as absorption;
\item[1.] likely false---though cannot name alternate metal-line
  identification, better data would probably confirm this as not
  \ion{C}{4} absorption;
\item[2.] likely true---though sparse supporting evidence (e.g.,
  associated lines), better data would probably confirm this as
  \ion{C}{4} absorption; or
\item[3.] definitely real---associated with other lines and\slash or
  shows clear correlations in line profiles.
\end{itemize}
\vspace{-0.5\baselineskip} A single author was trusted to accurately
assign ratings of 0 or 3; these doublets were not viewed again. The
more ambiguous cases (ratings 1 and 2) were reviewed by at least one
additional author until consensus was achieved. We grouped the
doublets into systems with $\dvabs < 250\kms$. 

Though we excluded the BAL QSOs from \citet{shenetal11}, we found
several sightlines with strong, self-blended, highly blueshifted
\ion{C}{4} absorbers. Since we focused on the {\it intergalactic}
absorption systems, we excluded these 138 sightlines from further
analysis and labeled them ``visual BAL'' QSOs in Table \ref{tab.los}
(binary flag $f_{\rm BAL}=12$). In addition, the extremely strong
absorption lines in these spectra corrupted the automated continuum
fitting algorithm. 

The final catalog includes all doublets with ${\rm rating} \ge 2$.
From the initial 29,789 candidates, 743 were in sightlines excluded as
visual BAL QSOs. In the remaining 29,046 candidates, we found 16,459
real \ion{C}{4} doublets (see Table \ref{tab.civ}). A sample of
doublets are shown in Figure \ref{fig.civexmpl}.  Ultimately, we
analyzed the 14,772 with $\dvqso < -5000\kms$ (see
\S\ref{subsec.freqdistr}).


We assessed the effects of blending and bad continua by inspecting
4500 doublets chosen at random. We flagged doublets where the 1548,
1550, or both lines were blended or otherwise problematic (e.g., the
bounds were unrealistic) and where the continua ought to be adjusted
locally (e.g., due to strong absorption or neighboring emission
lines). The majority of our analysis depended on \EWlin{1548}, so it
mattered most what fraction of the 1548 lines were troublesome. 

Over 90\% of the random sample had ``perfect'' treatment of the 1548
line, and over 95\% had acceptable continuum fits. The problematic
doublets showed some redshift dependence due to the confused nature in
the sky line region $\lambda_{\rm obs } \gtrsim 7500\Ang$ or $\zciv
\gtrsim 3.8$, where the fraction of problematic 1548 lines jumped from
$\approx10\%$ to $\approx20\%$.  The stronger lines ($\EWlin{1548}
\gtrsim 2\Ang$) were more often blended. These systems tended to be
self-blended and would generally be problematic to surveys using
boxcar summation for equivalent widths, such as ours. To maintain
the largely automated, and hence objective and repeatable, nature of
the survey, we decided to accept the imperfections and assess the
strength of our results in light of them.

\begin{deluxetable*}{rllr@{\,$\pm$\,}lr@{\,$\pm$\,}lr@{\,$\pm$\,}lr@{\,$\pm$\,}lc}
\tablewidth{0pc}
\tablecaption{\ion{C}{4} System Summary\label{tab.civ}}
\tabletypesize{\scriptsize}
\tablehead{ 
\colhead{(1)} & \colhead{(2)} & \colhead{(3)} & 
\multicolumn{2}{c}{(4)} & \multicolumn{2}{c}{(5)} & 
\multicolumn{2}{c}{(6)} & \multicolumn{2}{c}{(7)} & \colhead{(8)} \\ 
\colhead{QSO ID} & \colhead{\zqso} & 
\colhead{\zciv} & 
\multicolumn{2}{c}{\EWlin{1548}} & \multicolumn{2}{c}{\EWlin{1550}} & 
\multicolumn{2}{c}{$C(\EWlin{1548})$} & \multicolumn{2}{c}{\logCIV} & \colhead{$f_{N}$} \\ 
 & & & 
\multicolumn{2}{c}{(\AA)} & \multicolumn{2}{c}{(\AA)} & \multicolumn{2}{c}{} & 
\multicolumn{2}{c}{($\log({\rm cm}^{-2})$)} & 
} 
\startdata 
52143-0650-199 &  1.8449 & 1.52755 &  0.567 &  0.147 &  0.981 &  0.162 &  0.562 &  0.206 & $>14.30$ &  0.14 &  3 \\ 
52143-0650-178 &  2.6404 & 2.23461 &  0.425 &  0.059 &  0.303 &  0.058 &  0.363 &  0.092 & $>14.10$ &  0.06 &  3 \\ 
                         &           & 2.43947 &  1.050 &  0.078 &  0.653 &  0.072 &  0.860 &  0.023 & $>14.53$ &  0.03 &  3 \\ 
52902-1091-546 &  2.5551 & 2.34348 &  0.734 &  0.202 &  0.898 &  0.198 &  0.713 &  0.192 & $>14.75$ &  0.14 &  3 \\ 
                         &           & 2.42902 &  1.053 &  0.191 &  0.733 &  0.196 &  0.860 &  0.066 & $?14.63$ &  0.20 &  7 \\ 
51791-0387-093 &  1.8973 & 1.77015 &  0.388 &  0.097 &  0.479 &  0.098 &  0.305 &  0.132 & $>14.01$ &  0.13 &  3 \\ 
52235-0750-550 &  2.6383 & 2.16695 &  0.581 &  0.123 &  0.594 &  0.127 &  0.576 &  0.161 & $>14.23$ &  0.09 &  3 \\ 
54389-2822-423 &  2.7668 & 2.36223 &  0.738 &  0.087 &  0.332 &  0.088 &  0.716 &  0.073 & $ 14.34$ &  0.06 &  1 \\ 
54327-2630-423 &  2.7251 & 2.23946 &  0.489 &  0.100 &  0.390 &  0.102 &  0.458 &  0.151 & $>14.16$ &  0.11 &  3 \\ 
54452-2824-554 &  2.5566 & 2.08349 &  0.837 &  0.040 &  1.050 &  0.042 &  0.781 &  0.025 & $>14.55$ &  0.02 &  3 \\ 
                         &           & 2.08788 &  0.971 &  0.039 &  0.398 &  0.039 &  0.836 &  0.014 & $ 14.46$ &  0.02 &  1 \\ 
                         &           & 2.14618 &  0.769 &  0.041 &  0.462 &  0.037 &  0.738 &  0.030 & $>14.39$ &  0.03 &  3 \\ 
                         &           & 2.21361 &  0.381 &  0.039 &  0.254 &  0.041 &  0.294 &  0.059 & $>14.00$ &  0.05 &  3 \\ 
                         &           & 2.38185 &  0.395 &  0.039 &  0.284 &  0.039 &  0.317 &  0.062 & $>14.05$ &  0.05 &  3 \\ 
52143-0650-561 &  1.7593 & 1.63718 &  0.393 &  0.049 &  0.185 &  0.053 &  0.314 &  0.075 & $ 13.99$ &  0.07 &  1 \\ 
52991-1489-581 &  1.8536 & 1.69764 &  1.013 &  0.090 &  0.812 &  0.088 &  0.849 &  0.030 & $>14.55$ &  0.04 &  3 \\ 
52203-0685-567 &  2.0921 & 1.70301 &  1.165 &  0.186 &  0.669 &  0.180 &  0.877 &  0.038 & $ 14.57$ &  0.08 &  1 \\ 
                         &           & 1.75102 &  0.875 &  0.121 &  0.827 &  0.122 &  0.799 &  0.070 & $>14.54$ &  0.07 &  3
\enddata 
\tablecomments{ 
For each sightline (identified in Columns 1 and 2), every confirmed doublet is listed by the redshift of its C\,IV 1548 line (Column 3).
The rest equivalent widths of the C\,IV lines are given in Columns 4 and 5.
In Column 6, we give the completeness fraction for the doublet from the whole survey average.
The C\,IV column density (Column 7) is the combined value from the AODM measurements in both lines.
The binary column density flag $f_{N}$ is composed of: 1 = good to analyze; 2 = lower limit; 4 = upper limit; 8 = un-weighted average; and 16 = default to line value with greater significance (Column 8). 
(This table is available in a machine-readable form in the online journal.
A portion is shown here for guidance regarding its form and content.)
} 
\end{deluxetable*} 


\section{Completeness Tests}\label{sec.complete}

We test our survey completeness by generating a library of synthetic
profiles, randomly distributing them in a subset of the sightlines,
and tracking which simulated doublets were recovered.  The goal was to
populate a grid of doublet redshifts and rest equivalent widths with
the fraction of recovered and accepted simulated doublets in each grid
cell. We aimed to sample the full distribution of completeness limits
so that we could leverage the large number of sightlines; for example,
two spectra with 50\% completeness in one $(\EWr,\zciv)$ cell are
equivalent to one sightline with 100\% completeness for the same
parameters \citep{matejekandsimcoe12}. In this section, we detail
how we generated synthetic profiles, input and recover them, and
measure the completeness, in addition to discussing biases.

\subsection{Simulated Doublets\label{subsec.civprof}}

We parameterized the simulated doublets to produce absorption lines
that look like the diversity of profiles observed in the SDSS spectra.
We created a Voigt profile library with a uniform distribution of
equivalent widths, since we aimed to generate uniform errors on
detection completeness as a function of \EWlin{1548}. Each
component had a column density that was sampled linearly from $11 <
\logCIV < 14$ and a Doppler parameter randomly selected in the range
$5\kms < b < 40\kms$. For each system, one primary component was
generated, and a random draw from a Poisson distribution, with mean of
$\mu^{\rm Poiss}_{\rm comp}$, determined the number of additional
components for the system. The velocity offset for each component was
drawn from a Gaussian distribution with standard deviation
$\sigma^{\rm Gauss}_{\delta v}$, clipped to keep all components within
the maximum offset $\delta v_{\rm max}$ relative to the system
redshift.

The bulk of our profiles were generated with $\mu^{\rm Poiss}_{\rm
  comp} = 4$ or 7, $\sigma^{\rm Gauss}_{\delta v} = 100\kms$, and
$\delta v_{\rm max} = 300\kms$, which were roughly based on the
component properties measured by \citet{boksenbergetal03ph}. However,
to reproduce the diversity of profiles found in the SDSS data, we
had to modify the parameters (e.g., larger $\mu^{\rm Poiss}_{\rm
  comp}$ and $\delta v_{\rm max}$), and we limited the alternate
parameters to the larger equivalent width regime (e.g., $\EWr \ge
1.2\Ang$), where \citet{boksenbergetal03ph} did not have many systems.

We generated over $10^{5}$ Voigt profiles. Then we randomly selected
or duplicated (in extreme equivalent width regimes) profiles to
uniformly sample $0.05\Ang < \EWlin{1548} < 3.25\Ang$ in bins of
$0.05\Ang$, resulting in a library of 32,000 profiles.

\begin{figure*}[hbt]
  \begin{center}
    \includegraphics[width=0.94\textwidth]{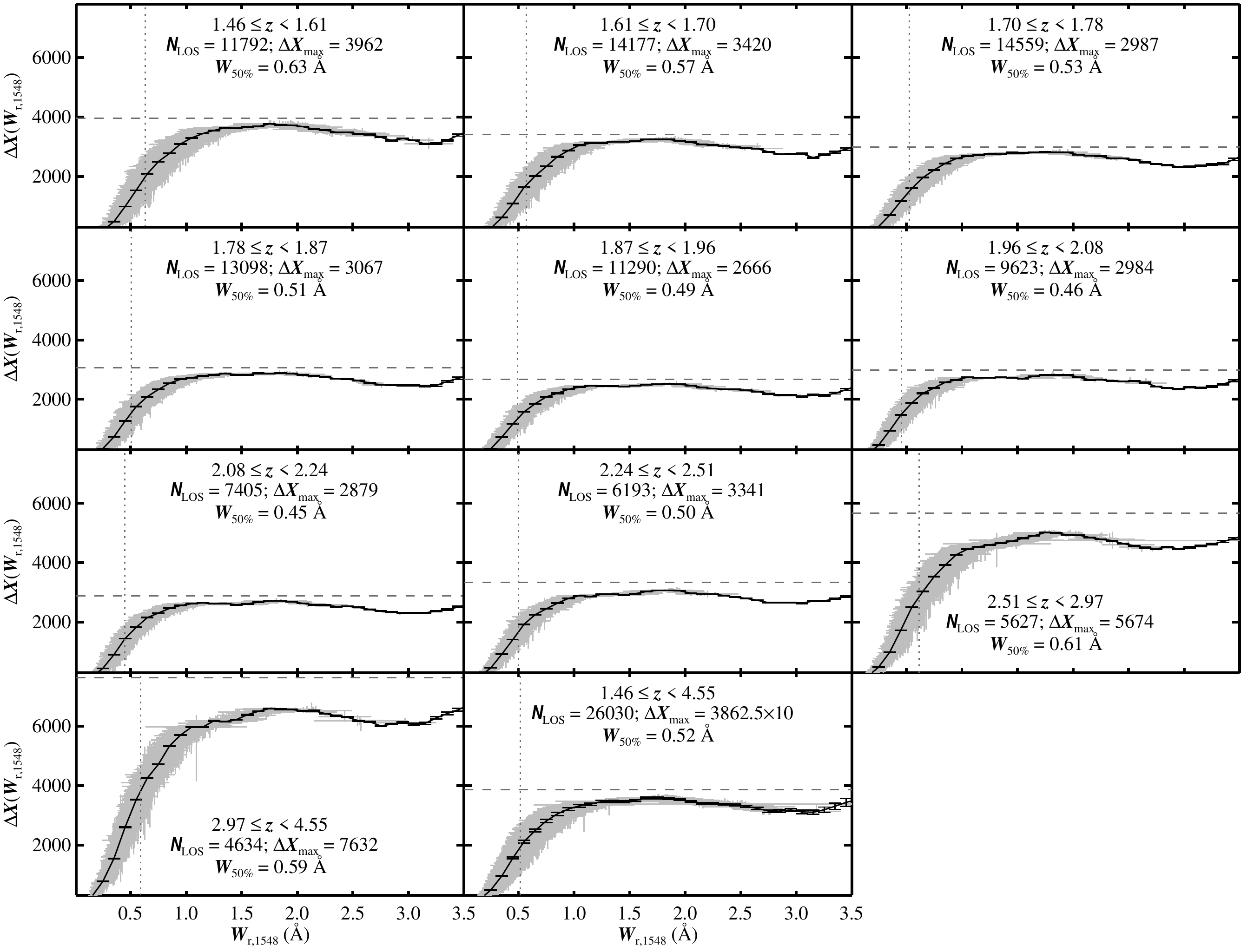}
  \end{center}
  \caption[Results from Monte-Carlo completeness tests.]
  {Results from Monte-Carlo completeness tests. The
    completeness-corrected co-moving path length as a function of rest
    equivalent width \DX{\EWlin{1548}}\ is shown for the 11 redshift
    bins used in the current study. The black curve and errors are the
    completeness curves (Equation \ref{eqn.dxwgrid}), and the gray
    points and errors are the observations (Equation
    \ref{eqn.dxw}). The horizontal, dashed line traces the maximum
    path length available in the bin, and the vertical, dotted line
    indicates the equivalent width where we are 50\% complete.
    \label{fig.cmplt}
  }
\end{figure*}

\subsection{Monte Carlo Procedures\label{subsec.cmpltproc}}

We have two Monte Carlo completeness tests, referred to as basic and
user.  The basic test included all of the automated procedures, from
continuum fitting to candidate selection. It was applied to the
largest fraction of sightlines ($\approx 30\%$). We tested the user
bias by visually inspecting the candidates in $\approx6\%$ of the
sightlines. The subsamples were set by requiring at least one
sightline in each bin of $\Delta \zqso = 0.25$ for $\zqso \ge 1.7$ and
$\Delta \langle {\rm S/N} \rangle = 0.25\,{\rm pixel}^{-1}$ for $\langle
{\rm S/N} \rangle \ge 4\,{\rm pixel}^{-1}$. For the basic test, with
more sampled sightlines, we kept sampling from the sightlines in these
$\Delta \zqso$ and $\Delta \langle {\rm S/N} \rangle$ bins, so that
the bins with more sightlines had more tested for completeness. Extra
high-redshift sightlines were included in the user test in increase
the statistics.

We chose to input the simulated doublets into the actual spectra in
order to sample realistic data. We ``cleaned'' each spectrum by
removing a random 30\% of the automatically detected absorption
features. The flux between the wavelength bounds (see
\S\ref{subsec.meas}) was replaced by continuum with a scatter drawn
from neighboring pixels, and the new error reflected those same
pixels. By leaving the remaining absorption features in, we were able
to measure the effects of blending and of absorption lines on the
automated continuum fit and line-finding algorithm.

We set a fiducial absorber redshift density $\dNCIVdz=5$ to determine
how many simulated doublets to input into the cleaned spectrum. This
translated to typically injecting one to three doublets per loop,
and the test iterated until at least 1000 simulated doublets were
input per sightline. We did not change the error array when injecting
profiles, which slightly but systematically lowered the signal-to-noise
ratio of the simulated doublets.

For every sightline in the basic test, a random sample of profiles
were drawn from our library and assigned redshifts to fall within the
limits of the current sightline. The spectrum was ``cleaned'' to start
and every 50th loop thereafter. We injected one to three simulated
doublets, fit the hybrid-continuum (see \S\ref{subsec.conti}), and ran
the automated candidate identification algorithm (see
\S\ref{subsec.cand}). Any injected profile was flagged as recovered if
the automatically identified candidate bounds spanned the observed
wavelength of the simulated system. For any profile that was {\it not}
recovered, we measured the flux-weighted redshift, equivalent width,
and AOD column density at the expected location. The input and
recovered information was stored for later processing (see
\S\ref{subsec.cmpltcorr}).

The user completeness test served a dual purpose. It tested the
effects of human bias (e.g., we were less likely to accept real doublets at
high redshift due to poor sky subtraction at $\gtrsim 7000\Ang$) and
our accepted false-positive rate (i.e., how often we rated non-\ion{C}{4}
lines as \ion{C}{4} absorption). The steps were largely the same for
one sightline in the user test as in the basic, but we ``cleaned'' the
profile every iteration and an author rated the automatically detected
candidates. However, we modified the simulated doublets to have an
unphysical characteristic wavelength separation of 4.8\Ang\ or
924\kms\ so that any candidate was most likely either simulated or
spurious. The fake $\lambda\lambda1547,1552$ profile library was
truncated to $\EWlin{1547} \le 1\Ang$. We did not simulate other
absorption lines because, fundamentally, we are conducting a blind
\ion{C}{4} survey.

\begin{figure*}[hbt]
  \begin{center}$
    \begin{array}{cc}
  \includegraphics[width=0.47\textwidth]{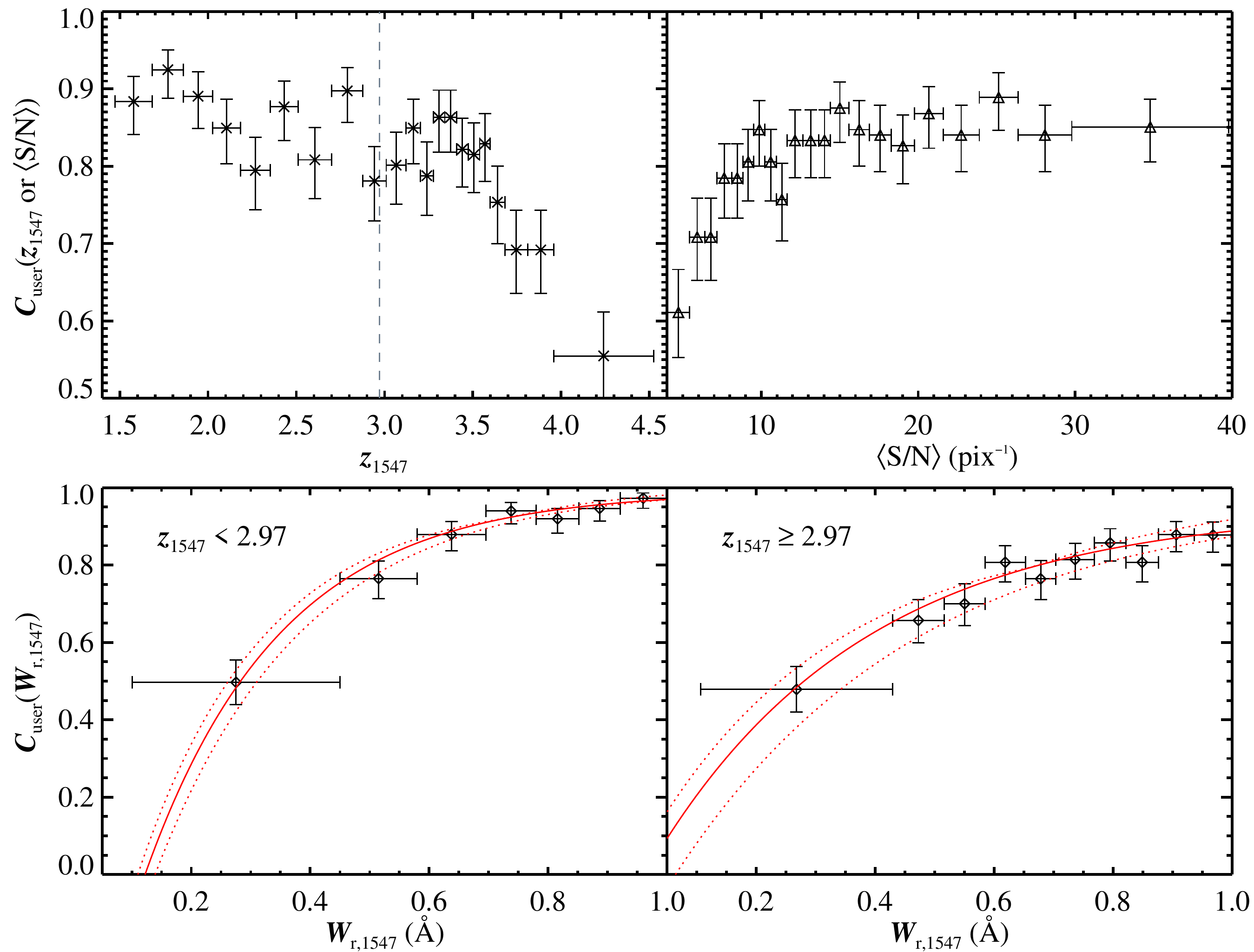}  & 
  \includegraphics[width=0.47\textwidth]{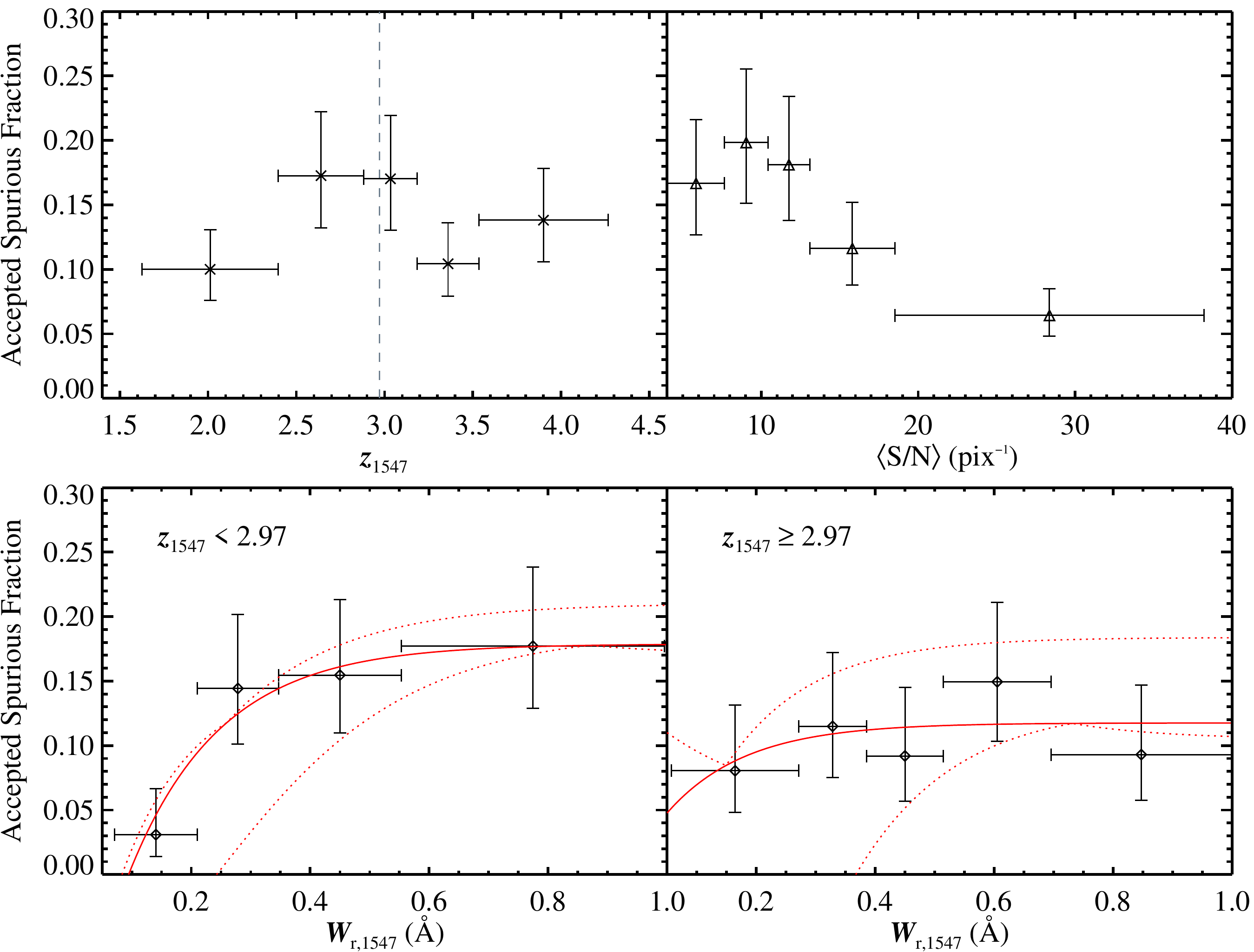} 
  \end{array}
 $\end{center}
  \caption[Biases of visual verification and trends of spurious detections.]
  {Biases of visual verification and trends of spurious
    detections. {\it Left}: We tested the ``user bias'' by rating fake
    $\lambda\lambda1547,1552$ doublets that were injected and
    automatically recovered as candidates. The four panels show the
    completeness of our ability to correctly rate true doublets, as a
    function of redshift, spectrum $\langle {\rm S/N} \rangle$, and
    equivalent width in two redshift bins. The redshift cut (vertical
    line) was determined by the decrease in $C_{\rm user}(z_{1547})$
    due to the sky-line region at $\approx 7000\Ang$ and by the
    highest redshift SDSS bin. The (red) solid and dashed lines are
    the best-fit model (see Equation \ref{eqn.czwuserfit}) and its
    1-$\sigma$ errors. {\it Right}: At the same time, we estimated the
    accepted false-positive fraction as functions of redshift,
    signal-to-noise, and equivalent width. These spurious pairs of
    lines were {\it not} injected but were automatically
    recovered. The lower panels show the best-fit model and errors to
    the accepted spurious fraction over \EWlin{1547}.
    \label{fig.userbias}
  } 
\end{figure*}

\subsection{Completeness Correction and Unblocked Co-Moving
  Path Length}\label{subsec.cmpltcorr} 

We combined the completeness tests in bins of redshift and onto grids
of equivalent width so that we correct any detected absorber based on
its completeness fraction.  The estimated completeness fraction for a
given grid point $(z_{\rm g},W_{\rm g})$ is simply $\Num_{\rm
  accept}/\Num_{\rm input}$, the fraction of input \ion{C}{4} doublets
that are recovered automatically and accepted by the user, in the
given grid cell. The two completeness tests separately measured the
recovering and accepting effects so that the final completeness fraction
is estimated by:
\begin{equation}
  C(z_{\rm g},W_{\rm g}) = \frac{ \Num_{\rm rec}(z_{\rm g},W_{\rm g})
  } { \Num_{\rm input}(z_{\rm g},W_{\rm g}) }\frac{ \Num_{\rm accept}(z_{\rm g},W_{\rm g})
  } { \Num_{\rm rec}(z_{\rm g},W_{\rm g}) } {\rm ,} \label{eqn.czw}
\end{equation}
where the right-hand product can be thought of as $C_{\rm basic}C_{\rm
  user}$.  The completeness grid $C(z_{\rm g},W_{\rm g})$ collapses to
a curve $C(W_{\rm g})$ in a fixed redshift bin. The basic completeness
uncertainty $\sigma_{C_{\rm basic}}$ is estimated by the Wilson score
interval for a binomial distribution \citep{wilson27}. This confidence
interval estimator is well behaved for small $\Num_{\rm input}$
and\slash or for extreme completeness fractions.

The statistics on the user completeness were naturally smaller than
for the basic, so we fit $C_{\rm user}(\EWr)$ with the following model:
\begin{equation}
  C_{\rm user}(\EWr) = C_{0} \big(1-e^{\beta(\EWr - W_{0})} \big)  \label{eqn.czwuserfit}
\end{equation}
in two redshift bins. The dividing $z = 2.97$ corresponded to the
beginning of the sky line region at $\approx 7000\Ang$, with its
resulting decrease in $C_{\rm user}$ due to confusion, and matched the
start of the highest redshift bin. The fit uncertainties were
estimated from Monte Carlo re-sampling of the equivalent-width errors,
and the final error on $C(\EWr)$ was propagated from $\sigma_{C_{\rm
    basic}}$ and $\sigma_{C_{\rm user}}$. We extrapolated the fits to
larger equivalent widths when we calculated the full completeness
fraction. The user completeness test is discussed in detail in
\S\ref{subsec.biases} below.

We scaled $C(z_{\rm g},W_{\rm g})$ by the fraction of the
total path length {\it not} obscured by doublets with greater
equivalent widths, in the redshift bin. All \ion{C}{4} lines blocked
$<2\%$ of the total survey path length (see Table \ref{tab.los}). 

In actuality, $\Num_{\rm input}$ and $\Num_{\rm rec}$ contain the
information only for the profiles and sightlines actually sampled in
the basic completeness test. Any sightline that did not have
a measurement for one or more quantities was accounted for with the
average of the sightlines actually tested in the requested $\Delta
\zqso$ and $\Delta \langle {\rm S/N} \rangle$ bin.

The unblocked co-moving path length\footnote{The co-moving path length
  is related to redshift as follows: $X(z) = 2\sqrt{ \Omega_{\rm
      M}(1+z)^{3} + \Omega_{\Lambda}}/(3 \Omega_{\rm M})$.}  for each
grid cell is calculated by simply multiplying the completeness
fraction by the total path length available in each redshift bin:
\begin{eqnarray}
\DX{W_{\rm g}} & = & \DX{z_{\rm g}} C(z_{\rm g},W_{\rm
  g}) \label{eqn.dxwgrid} \\
\sigDX^2{}_{(W_{\rm g})} & = & \DX{z_{\rm g}}^2 \sigma_{C(z,W)}^2
{\rm .} \nonumber
\end{eqnarray}
In Figure \ref{fig.cmplt}, the black curves are the grid values
$\DX{W_{\rm g}}$ and errors.

The unblocked co-moving path length to which our survey is sensitive
for any given detected equivalent width $W_{\rm i}$ is interpolated
from the grid of $\DX{W_{\rm g}}$:
\begin{eqnarray}
\DX{W_{\rm i}} & = &  {\rm interpol}( \DX{W_{\rm g}},\ W_{\rm g},\ W_{\rm
i})  \label{eqn.dxw} \\
\sigDX^2 & = & {\rm interpol}( \sigDX^2{}_{(W_{\rm g})},\ W_{\rm g},\ W_{\rm
i}) + \nonumber \\
& & \bigg( \DX{W_{\rm i}} - \DX{W_{\rm i} \pm \sigma_{W_{\rm i}}} \bigg)^2 \nonumber
{\rm .}
\nonumber 
\end{eqnarray}
The error on $\DX{W_{\rm i}}$ accounts for the error in our
completeness correction (\sigDX) and for the uncertainty in our
equivalent width measurement ($\sigma_{W_{\rm i}}$). We plot each
doublet's \DX{\EWlin{1548}}\ in gray in Figure
\ref{fig.cmplt}. Clearly, the uncertainty in equivalent width
dominates the error in the completeness correction.

In all redshift bins, the typical completeness fraction reached 50\%
by $\EWlin{1548} \approx 0.6\Ang$, a value we frequently use as the
minimum in subsequent analyses.

\begin{figure*}[hbt]
  \begin{center}
  \includegraphics[width=0.94\textwidth]{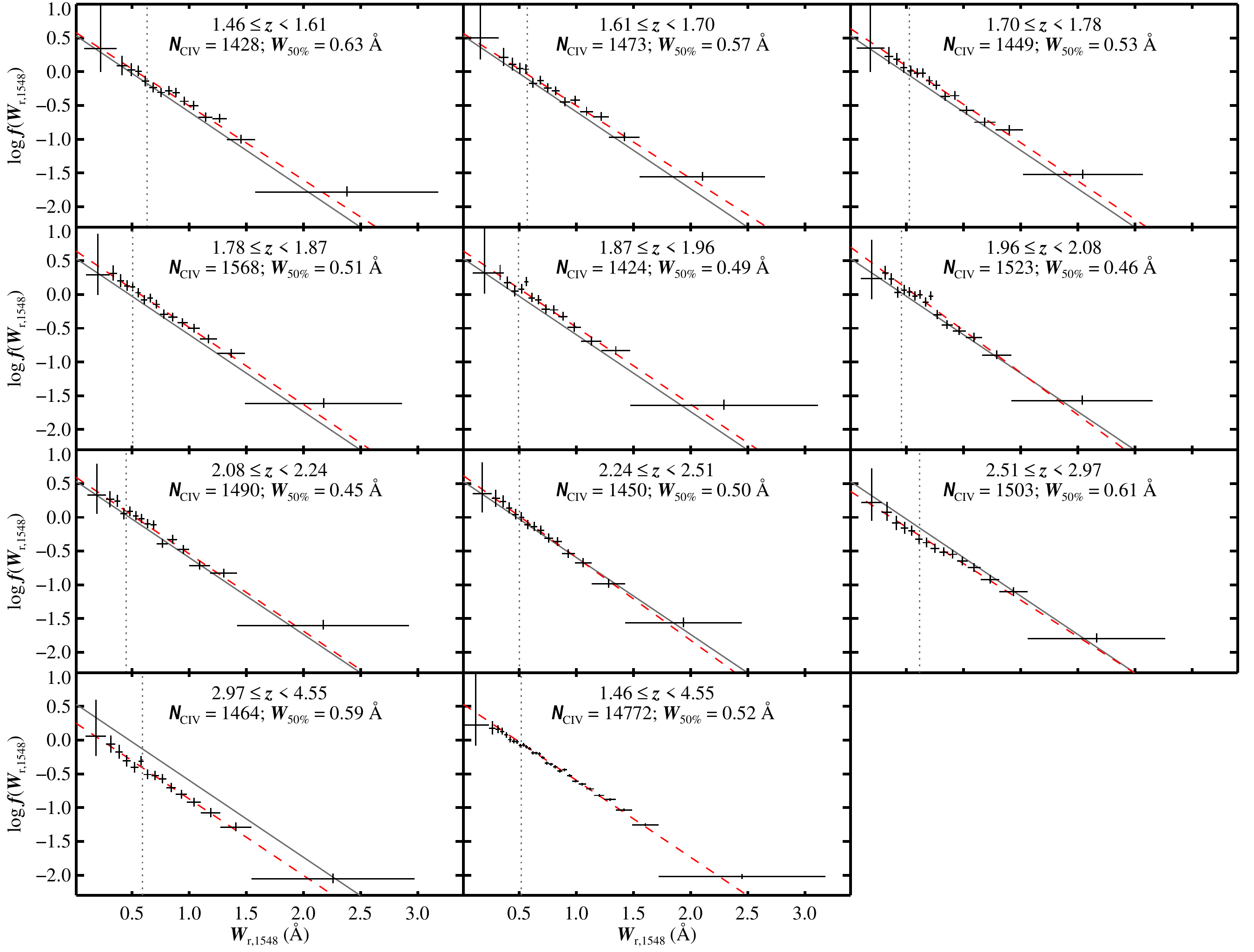} 
  \end{center}
  \caption[Equivalent width frequency distributions.]
  {Equivalent width frequency distributions. The maximum likelihood
    fits of an exponential function are the dashed (red) lines, for
    each redshift bin, and the solid (gray) line, for the full
    sample. There is little evolution with redshift. The observations
    have been completeness corrected, and the redshift-specific 50\%
    completeness limits are the vertical dotted lines.
    \label{fig.fw}
  }
\end{figure*}

\begin{figure}[hbt]
  \begin{center}
  \includegraphics[width=0.47\textwidth]{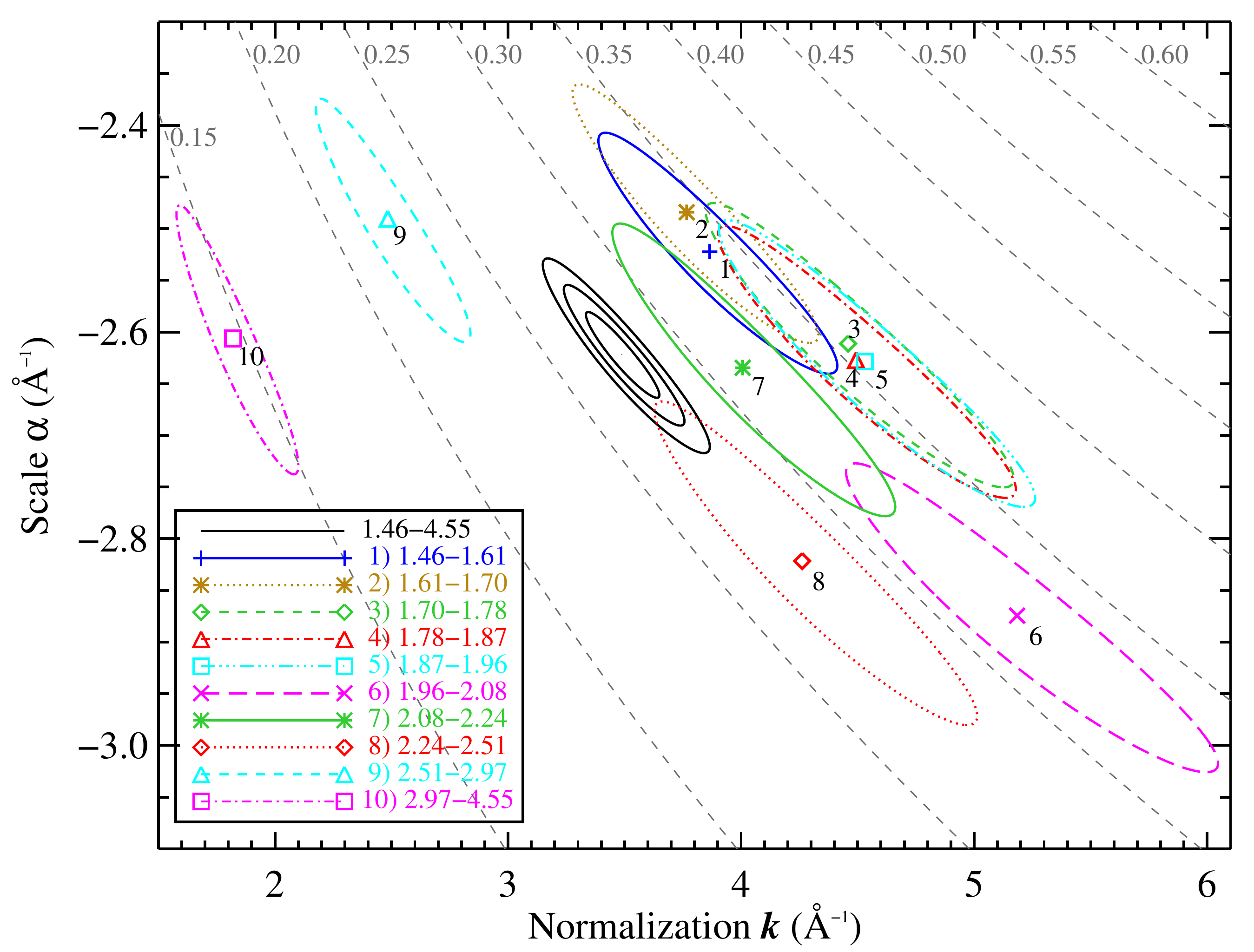} 
  \end{center}
  \caption[Best-fit \ff{\EWr} parameters and errors.]
  {Best-fit \ff{\EWr}\ parameters and errors. We fit the frequency
    distribution with an exponential function (see Equation
    \ref{eqn.fexp}). The best-fit normalization $k$ and scale $\alpha$
    and the 1-$\sigma$ error ellipses are plotted for the ten small redshift
    bins (numbered points); the 1, 2, and 3-$\sigma$ contours are shown
    for the fit to the full sample (black ellipses). The best-fit
    parameters smoothly evolve with redshift, as seen by comparing the
    ellipses with the constant \dNCIVdX\ curves (gray, dashed lines). 
   \label{fig.fwellipse}
  }
\end{figure}

\subsection{Automated and Interactive Biases}\label{subsec.biases} 

The completeness curves do not monotonically increase to 100\% at
large equivalent width (see Figure \ref{fig.cmplt}). Instead, they
roll off at $\EWlin{1548} \approx 2\Ang$. This feature results from
the incompleteness in the broad (self-blended) \ion{C}{4} search and,
to a much lesser extent ($< 5\%$), from broad profiles being over-fit by
the continuum algorithm, which led them to being missed in the
automated search.

The user completeness test measured our ability to correctly rate
real, automatically detected doublets (``true positives'') as well as
the rate at which we include spurious pairs of lines as true doublets
(``accepted false negatives'').  We injected $\Num_{\rm input} = 5021$
fake doublets with $\EWr < 1\Ang$ and $\dvqso < -5000\kms$, and
$\Num_{\rm rec} = 3070$ were automatically recovered. Of these, we
correctly rated $\Num_{\rm accept} = 2489$. The left panels of Figure
\ref{fig.userbias} show the trends of the user completeness with
redshift, spectrum $\langle {\rm S/N} \rangle$, and equivalent width.

Generally, $C_{\rm user}$ decreases with increasing redshift and
decreasing $\langle {\rm S/N} \rangle$. Sky lines become numerous at
$\lambda \gtrsim 7000\Ang$, and poor sky subtraction leaves features
in the spectrum that mimic absorption lines (see Figure
\ref{fig.conti}). In the visual verification step, a larger fraction
of real doublets are rejected due to the severe confusion. Lower
signal-to-noise spectra induces the same effect.

There were an additional 966 spurious candidates brought forth by the
automated search, of which we incorrectly accepted 121.  The accepted
false-positive fraction peaks at $\approx 20\%$ at $z \approx 3$ and
$\langle {\rm S/N} \rangle \approx 10$. However, our acceptance
fraction grows sharply at $\EWr \lesssim 0.6\Ang$ and plateaus to 18\%
and 12\% for $z < 2.97$ and $\ge 2.97$, respectively.  The estimated
co-moving line densities of accepted false positives with $\EWr \ge
0.6\Ang$ are $\dNFkIVdX = 0.022$ and 0.020 for the low and high
redshift bins, respectively.

We detail how we applied corrections for the accepted false-positive
rate in the next section.



\begin{deluxetable*}{cccccccccccc}
\tablewidth{0pc}
\tablecaption{\ion{C}{4} Results Summary \label{tab.freqdistr}}
\tabletypesize{\scriptsize}
\tablehead{ 
\colhead{(1)} & \colhead{(2)} & \colhead{(3)} & \colhead{(4)} & 
\colhead{(5)} & \colhead{(6)} & \colhead{(7)} & \colhead{(8)} & 
\colhead{(9)} & \colhead{(10)} & \colhead{(11)} & 
\colhead{(12)} \\ 
\colhead{$\langle z \rangle$} & \colhead{$z_{\rm lim}$} & 
\colhead{$\Num_{\rm obs}$} & \colhead{$\Delta X_{\rm max}$} & \colhead{$W_{50\%}$} & 
\colhead{\dNCIVdz} & \colhead{\dNCIVdX} & 
\colhead{\OmCIV} & 
\colhead{$\Num_{\rm fit}$} & \colhead{$k$} & \colhead{$\alpha$} &  
\colhead{$\chi^{2}_{\rm red}$} \\ 
 & & 
 & & \colhead{(\AA)} 
 & & & 
\colhead{($\times10^{-8}$)} & & 
\colhead{(\AA$^{-1}$)} & \colhead{(\AA$^{-1}$)} &  
}
\startdata
  1.96274 & $[  1.46623,   4.54334]$ &  14772 &  38624 &  0.52 & $ 0.92^{+ 0.02}_{- 0.01}$ & $ 0.275^{+ 0.004}_{- 0.004}$ & $ 1.71^{+ 0.20}_{- 0.20}$ &    8918 & $   3.49^{+   0.31}_{-   0.30}$ & $  -2.62^{+   0.04}_{-   0.04}$ &   0.577 \\ 
  1.55687 & $[  1.46623,   1.60986]$ &   1428 &   3962 &  0.63 & $ 0.95^{+ 0.03}_{- 0.03}$ & $ 0.332^{+ 0.011}_{- 0.010}$ & $ 2.18^{+ 0.39}_{- 0.39}$ &    1025 & $   3.87^{+   0.94}_{-   0.82}$ & $  -2.52^{+   0.11}_{-   0.12}$ &   1.560 \\ 
  1.65971 & $[  1.61004,   1.69993]$ &   1473 &   3420 &  0.57 & $ 1.00^{+ 0.04}_{- 0.03}$ & $ 0.336^{+ 0.012}_{- 0.011}$ & $ 2.10^{+ 0.32}_{- 0.32}$ &     956 & $   3.77^{+   0.95}_{-   0.82}$ & $  -2.48^{+   0.12}_{-   0.13}$ &   0.876 \\ 
  1.73999 & $[  1.70035,   1.78000]$ &   1449 &   2987 &  0.53 & $ 1.08^{+ 0.04}_{- 0.04}$ & $ 0.356^{+ 0.013}_{- 0.012}$ & $ 2.17^{+ 0.28}_{- 0.28}$ &     902 & $   4.46^{+   1.24}_{-   1.06}$ & $  -2.61^{+   0.14}_{-   0.14}$ &   1.672 \\ 
  1.82360 & $[  1.78007,   1.86985]$ &   1568 &   3067 &  0.51 & $ 1.09^{+ 0.04}_{- 0.04}$ & $ 0.351^{+ 0.012}_{- 0.011}$ & $ 2.11^{+ 0.23}_{- 0.24}$ &     932 & $   4.49^{+   1.20}_{-   1.04}$ & $  -2.63^{+   0.13}_{-   0.13}$ &   1.291 \\ 
  1.91460 & $[  1.87000,   1.95998]$ &   1424 &   2665 &  0.49 & $ 1.12^{+ 0.04}_{- 0.04}$ & $ 0.355^{+ 0.013}_{- 0.012}$ & $ 2.15^{+ 0.23}_{- 0.23}$ &     824 & $   4.53^{+   1.28}_{-   1.09}$ & $  -2.63^{+   0.14}_{-   0.14}$ &   1.832 \\ 
  2.01778 & $[  1.96002,   2.07997]$ &   1523 &   2983 &  0.46 & $ 1.04^{+ 0.04}_{- 0.04}$ & $ 0.322^{+ 0.012}_{- 0.011}$ & $ 1.86^{+ 0.17}_{- 0.17}$ &     852 & $   5.18^{+   1.76}_{-   1.51}$ & $  -2.87^{+   0.15}_{-   0.15}$ &   2.838 \\ 
  2.15320 & $[  2.08009,   2.23997]$ &   1490 &   2878 &  0.45 & $ 1.04^{+ 0.04}_{- 0.04}$ & $ 0.312^{+ 0.012}_{- 0.011}$ & $ 1.92^{+ 0.17}_{- 0.17}$ &     807 & $   4.01^{+   1.21}_{-   1.04}$ & $  -2.63^{+   0.14}_{-   0.14}$ &   3.267 \\ 
  2.35608 & $[  2.24015,   2.50914]$ &   1450 &   3341 &  0.50 & $ 0.96^{+ 0.04}_{- 0.03}$ & $ 0.276^{+ 0.011}_{- 0.010}$ & $ 1.61^{+ 0.18}_{- 0.18}$ &     775 & $   4.26^{+   1.62}_{-   1.37}$ & $  -2.82^{+   0.15}_{-   0.16}$ &   1.296 \\ 
  2.72298 & $[  2.51028,   2.96976]$ &   1503 &   5673 &  0.61 & $ 0.83^{+ 0.03}_{- 0.03}$ & $ 0.221^{+ 0.009}_{- 0.008}$ & $ 1.44^{+ 0.22}_{- 0.22}$ &     969 & $   2.48^{+   0.74}_{-   0.65}$ & $  -2.49^{+   0.12}_{-   0.12}$ &   0.837 \\ 
  3.25860 & $[  2.97005,   4.54334]$ &   1464 &   7632 &  0.59 & $ 0.59^{+ 0.02}_{- 0.02}$ & $ 0.145^{+ 0.006}_{- 0.005}$ & $ 0.87^{+ 0.13}_{- 0.13}$ &     876 & $   1.82^{+   0.75}_{-   0.65}$ & $  -2.61^{+   0.13}_{-   0.13}$ &   1.548
\enddata
\tablecomments{
Summary of the most common redshift bins and data used for the various analyses.
Columns 1--2 give the median, minimum, and maximum redshifts for the observed number of doublets (Column 3), and the maximum co-moving pathlength in the redshift bin is given in Column 4.
The 50\% completeness limit from the Monte Carlo tests is in Columns 5.
The redshift and co-moving absorber line densities for $\EWr \ge 0.6\Ang$ are in Columns 6--7.
In Column 8, the \OmCIV\ from summing the mass in the $\EWr \ge 0.6\Ang$ absorbers is a lower limit, since the majority of absorbers are saturated.
The frequency distribution was fit with an exponential $\ff{\EWr} = k \exp(\alpha\EWr)$ for $\Num_{\rm fit}$ absorbers with $\EWr \ge 0.6\Ang$ (Column 9), and the best-fit parameters are given in Columns 10--11.
The reduced $\chi^{2}$ from the best fit and \ff{\EWr}\ (in bins with $\approx100$ doublets each) is given in Column 12.
}
\end{deluxetable*}

\section{Results}\label{sec.results}

The bulk of the analysis of the \ion{C}{4} sample was performed either
on the whole dataset or in bins in redshift space.  The bins were
determined empirically from the $\dvqso < -5000\kms$ sample to have
about 1500 doublets per bin, and the \dvqso\ cut is explained
below.

\begin{figure}[hbt]
  \begin{center}
  \includegraphics[width=0.47\textwidth]{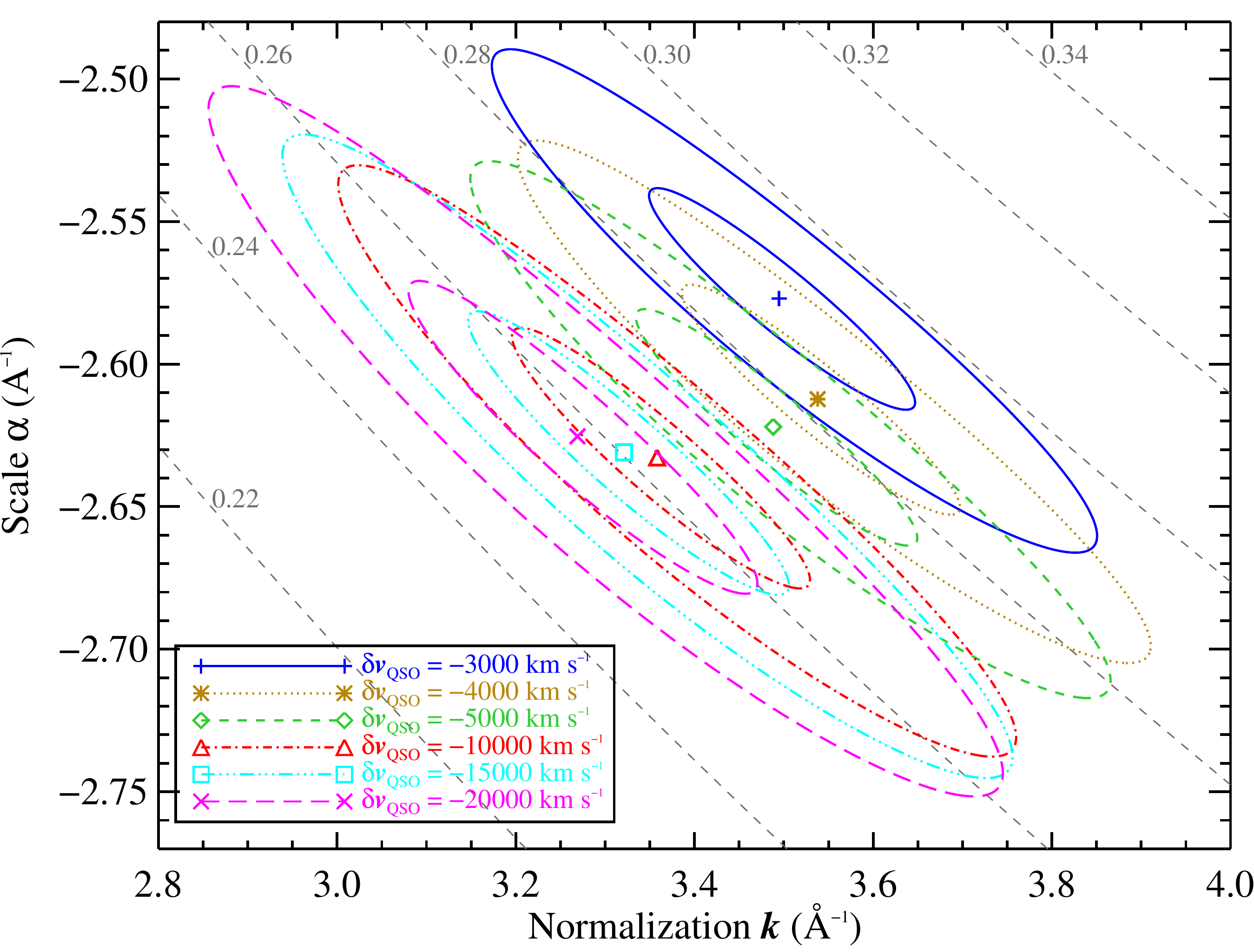} 
  \end{center}
  \caption[Best-fit \ff{\EWr} parameters for different dvqso\ values.]
  {Best-fit \ff{\EWr}\ parameters for different \dvqso\ values. We
    tested the effects of the \dvqso\ cut on the fit to the full
    \ion{C}{4} sample. The contours are 1 and $3\sigma$, and the gray,
    dashed lines are constant \dNCIVdX\ curves. We adopt
    $\dvqso=-5000\kms$ for the main analyses.
   \label{fig.fwdvqso}
  }
\end{figure}

\subsection{Frequency Distribution}\label{subsec.freqdistr}

The equivalent width frequency distribution \ff{\EWr}\ is the number
of detections $\Num_{\rm obs}(\EWr)$ per rest
equivalent width bin $\Delta \EWr$ per the total co-moving path length
available, in the given equivalent width bin, \DX{\EWr}:
\begin{eqnarray}
  \ff{\EWr} & = & \frac{\Num_{\rm obs}(\EWr)} {\Delta \EWr
    \,\DX{\EWr}}  \label{eqn.fewdef} \\
\sigma_{\ff{\EWr}}^2 & = & \ff{\EWr}^2 \Bigg( \bigg(
 \frac{\sigma_{\Num_{\rm obs}}}{\Num_{\rm obs}(\EWr)} \bigg)^2 + 
 \bigg( \frac{\sigDX}{\DX{\EWr}} \bigg)^2 
\Bigg) {\rm .} \nonumber
\end{eqnarray}
The error on $\Num_{\rm obs}$ is estimated from a Poisson distribution
if $\Num_{\rm obs} < 120$ and from a Gaussian approximation if
$\Num_{\rm obs} \ge 120$.  For \DX{\EWr}\ and \sigDX, we used
Equation \ref{eqn.dxw} with the center of the equivalent width bin
being $W_{\rm i}$ and $\sigma_{W_{\rm i}} = 0.5\Delta \EWr$. 

We used the maximum likelihood analysis of \citet{cookseyetal10} to fit
\ff{\EWr}\ with an exponential:
\begin{equation}
\ff{\EWr} = k e^{\alpha \EWr} \label{eqn.fexp} 
\end{equation}
(see Figure \ref{fig.fw}). The normalization $k$ and scale $\alpha$
were simultaneously fit, and the errors were estimated by the maximum
extent of the 1-$\sigma$ error ellipse on the likelihood surface (see
Figure \ref{fig.fwellipse}). All frequency distributions were fit over
the range $0.6\Ang \le \EWlin{1548} \le {\rm
  max}[\EWlin{1548}+\sigEWr]$.  The results were not strongly dependent
on the choice of the upper limit. The exponential model is a very good
description of the data, and the best-fit parameters are given in
Table \ref{tab.freqdistr}.


The best-fit parameters show smooth redshift evolution with respect to
the constant \dNCIVdX\ curves in Figure \ref{fig.fwellipse}. The
co-moving line density is simply the integral of the frequency
distribution from some limiting equivalent width $W_{\lim}$ to
infinity, and substituting in our exponential model, we see:
\begin{equation}
\frac{\displaystyle \ud \Num_{\Cthr}}{\displaystyle \ud X}\bigg |_{\rm fit} =
\frac{\displaystyle -k}{\displaystyle \alpha} e^{\alpha W_{\rm
    lim}} \label{eqn.dndxfit} {\rm .}
\end{equation}
By fixing \dNCIVdX, we can solve for the required normalization $k$
for any given $\alpha$. The 1-$\sigma$ ellipses are elongated in the
direction of the constant \dNCIVdX\ curves, and redshift evolution can
be seen by tracking systematic change perpendicular to these curves.
The lowest five redshift bins ($1.46 \le z < 1.96$) fall along roughly
the same constant \dNCIVdX\ curve. The next three highest redshift
bins ($1.96 \le z < 2.51$) have slightly smaller line densities. Then
there is almost a factor of two drop over the highest two redshift
bins ($2.51 \le z < 4.55$).

Since the accepted false-positive rate is essentially constant at
$\EWr \ge 0.6\Ang$ (see \S\ref{subsec.biases}), the frequency
distribution of accepted false positives is a scaled-down version of
the measured frequency distribution.  Therefore, we scale the original
$f_{0}(\EWr)$ as follows:
\begin{equation}
\ff{\EWr} = \bigg(1 - \frac{\displaystyle \dNFkIVdX}{\displaystyle
  \dNCIVdX} \bigg) f_{0}(\EWr) \label{eqn.afpfw}
\end{equation}
and propagate the errors. This results in a decrease of $\approx 6\%$
to 12\%, depending on the redshift.  For the exponential fits, we
scale the best-fit normalization $k_{0}$ in a similar fashion:
\begin{equation}
k = \bigg(1 - \frac{\displaystyle \dNFkIVdX}{\displaystyle
  (\dNCIVdX)_{\rm fit}} \bigg) k_{0} {\rm ,} \label{eqn.afpfwfit}
\end{equation}
but the denominator is the integrated line density from the
exponential model (Equation \ref{eqn.dndxfit}). We report the
propagated errors in Table \ref{tab.freqdistr} and in the text.
However, since we cannot compute the change in the likelihood surface,
we only shift the ellipses in Figures \ref{fig.fwellipse} and
\ref{fig.fwdvqso}.

\subsubsection{Effect of Blending and ``Intrinsic'' Absorbers}

As mentioned in \S\ref{subsec.verify}, blended profiles are an issue
for this survey, which relies on automated procedures and boxcar
summation to measure equivalent widths. We ran 1000 Monte Carlo
simulations to estimate the net effect of blending. There were over
$10^{6}$ (17\%) simulated doublets recovered in the completeness tests
that had measured equivalent widths more than $3\sigma$ larger than
the input value, indicative of blending. We measured the median
($\langle W_{\rm blend} \rangle = 0.2\Ang$) and standard deviation
($\sigma_{W_{\rm blend}} = 0.5\Ang$) of the distribution, where
$W_{\rm blend} = \EWlin{rec} - (\EWlin{input} +
3\sigma_{\EWlin{rec}})$.  For each realization, we tested the
worst-case scenario by decreasing the equivalent width of a random
20\% of the absorbers, which was the largest blended fraction
estimated in \S\ref{subsec.verify}. The magnitude of the decrease was
drawn randomly from the half of a Gaussian distribution below its
mean, set to $\langle W_{\rm blend} \rangle$, with standard deviation
$\sigma_{W_{\rm blend}}$. Then, the new sample was fit with an
exponential \ff{\EWr} model. The median best-fit parameters from these
Monte Carlo simulations were in very good agreement with those in
Table \ref{tab.freqdistr}, and the standard deviation of the Monte
Carlo results were ten-times smaller than the quoted
uncertainties. These simulations assessed the worst-case scenario, by
taking the largest estimated fraction of blended lines (see
\S\ref{subsec.verify}) and only decreasing the equivalent widths (as
opposed to also re-sampling the the other 80\% of the doublets).

In addition, the completeness corrections account for blending
since we compiled the curves in Figure \ref{fig.cmplt} with respect to
the {\it measured}---as opposed to input---equivalent widths. There
were other absorption lines in the spectra, and the randomly placed
simulated profiles were blended at a realistic rate, as seen by the
agreement of the blended fraction from our visual estimate
(\S\ref{subsec.verify}) and from the $10^{6}$ simulated doublets
described above. Thus, our results are robust to the effects of
blending.

We tested the effect of the \dvqso\ cut on our results by increasing
\dvqso\ and re-fitting the full redshift sample (see Figure
\ref{fig.fwdvqso}). Systems close to the background QSO could
potentially be high velocity, intrinsic absorbers and\slash or
affected by quasar clustering, enrichment, and\slash or local
ionization. The general trend for $\dvqso < -4000\kms$ is decreasing
normalization (i.e., $k$) with nearly constant shape ($\alpha$). The
decrease in $k$ is due to the number of absorbers decreasing faster
than the path length up to at least $\dvqso = -15000\kms$. For example,
with $\dvqso=-10000\kms$, we have 17.5\% less path length but 25\%
fewer $\EWr \ge 0.6\Ang$ doublets. This trend is partially explained
by quasars residing in dense environments
\citep[e.g.,][]{prochaskaetal11a}, where the chance of intersecting a
metal-enriched galaxy halo is increased.

However, this effect would not dominate out to $\dvqso =
-10000\kms$. The lack of ``convergence'' at very large \dvqso\ may be
partially due to poorly measured quasar
redshifts. \citet{hewettandwild10} re-measured the majority of the
redshifts for the \citet{schneideretal10} DR7 QSOs, with better
automated routines.
Adopting the \citet{hewettandwild10} redshifts affects over 80\% of
the total 16,459 \ion{C}{4} systems.  The majority of the quasar
redshifts {\it increased}, with a median $\langle \dvem \rangle =
381\kms$, standard deviation $764\kms$, and maximal extent $-4200\kms
\lesssim \dvem \lesssim +5700\kms$. In addition, there are
redshift-dependent fluctuations in \dvem. Using the
\citet{hewettandwild10} redshifts would affect our analysis; however,
the change in the total sample size, given a \dvqso\ cut, is less than
2\%. Given this small fraction and the incompleteness of the new
redshifts, we chose to continue with the \citet{schneideretal10}
redshifts, but we adopt $\dvqso=-5000\kms$ for the bulk of our
analyses which reduces the sample to 14,772 doublets.

\begin{figure}[hbt]
  \begin{center}
  \includegraphics[width=0.47\textwidth]{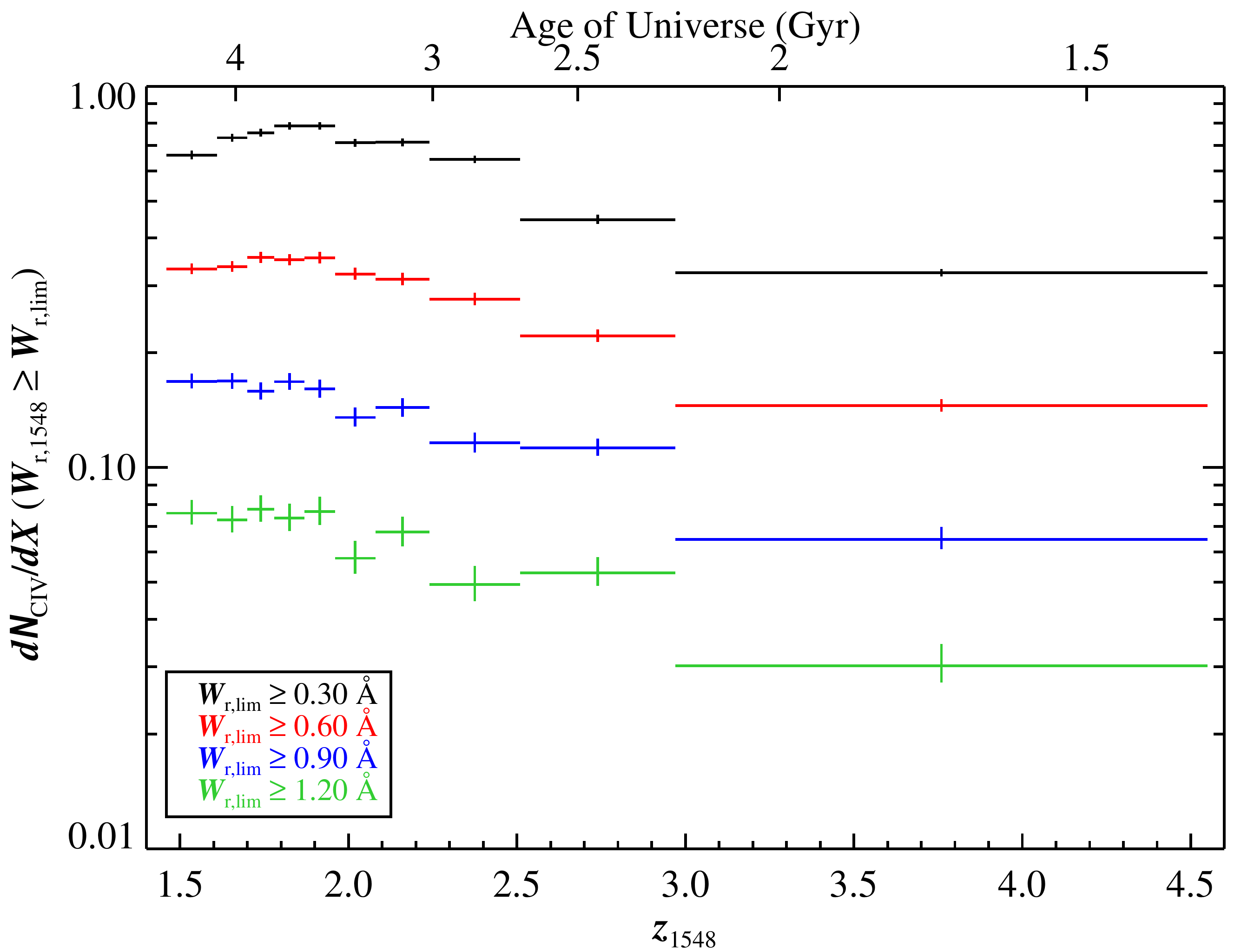} 
  \end{center}
  \caption[Co-moving \ion{C}{4} line density evolution.]
  {Co-moving \ion{C}{4} line density evolution. The number of
    absorbers per co-moving path length increases steadily and
    consistently from $\zciv = 4.5 \rightarrow\ \approx\!1.74$.  As
    expected from the nearly unchanging nature in the shape of
    \ff{\EWr}\ (see Figure \ref{fig.fw}), there is little dependence
    on \EWlin{\rm lim}, noting that we are typically 50\% at $\EWr
    \approx 0.6\Ang$.
    \label{fig.dndx}
  }
\end{figure}

\begin{figure*}[hbt]
 \begin{center}$
    \begin{array}{cc}
  \includegraphics[width=0.47\textwidth]{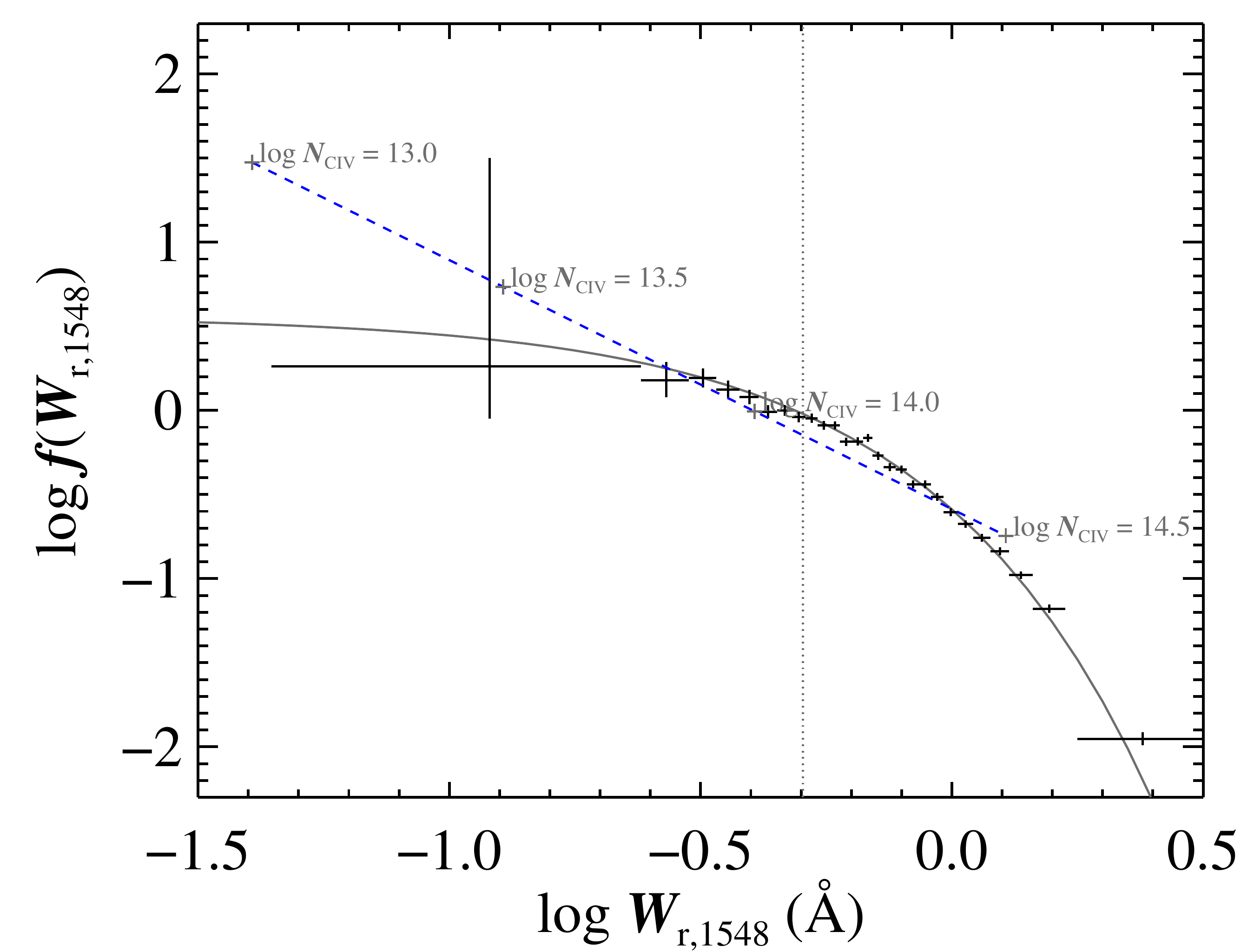}  & 
  \includegraphics[width=0.47\textwidth]{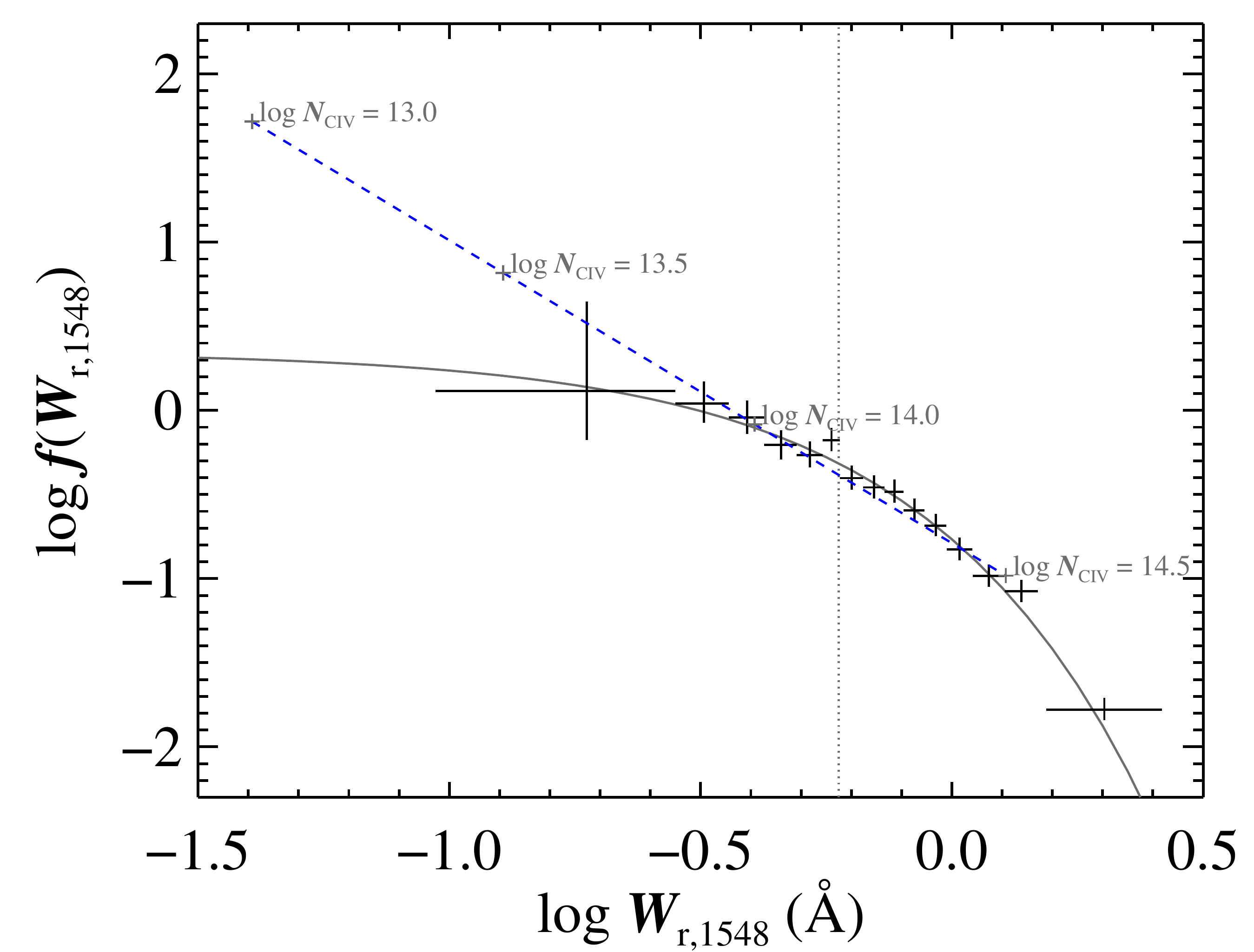} 
  \end{array}
 $\end{center}
 \caption[Comparing \ff{\EWr}\ to smaller, higher resolution studies.]
 {Comparing \ff{\EWr}\ to smaller, higher resolution studies. We show
   the best power-law fit from {\it left}: $1.6 \le z \le 3.6$
   \citep[$\alpha_{\rm pl}=-1.71\pm0.07$;][]{dodoricoetal10} and {\it
     right}: $2.9 \le z \le 3.54$ \citep[$\alpha_{\rm
     pl}=-1.8\pm0.1$;][]{songaila01} as the (blue) dashed lines. Our
   best exponential fit for each redshift bin is shown on top of the
   observed \ff{\EWr}\ (in bins of $\approx500$ and 100,
   respectively).  The vertical 50\%-complete line also happens to
   be around where a single \ion{C}{4} {\it component} saturates. The
   gray plus-signs indicate the equivalent widths where the labeled
   column densities fall.
    \label{fig.fwlit}
  }
\end{figure*}

\subsection{\ion{C}{4} Absorber Line Density}\label{subsec.dndx} 

We directly measured the absorber line density for doublets with $\EWr \ge
W_{\rm lim}$ as follows:
\begin{eqnarray}
\frac{ \ud \Num_{\mathrm{C\,IV}}}
{ \ud X} (\EWr \ge W_{\rm lim}) & = & \frac{\displaystyle
  \Num_{\rm C}(\EWr) }{\displaystyle \DX{z}} 
\label{eqn.dndx_sum}  \\
\sigma_{\ud \Num/\ud X}^{2} & = &  \bigg( \frac{ \sigma_{\Num_{\rm C}{} } }
 { \DX{z}}  \bigg)^2 
{\rm .} \nonumber 
\end{eqnarray}
The completeness-corrected number of absorbers in any given bin is the
completeness-weighted sum of the observed absorbers in the bin:
\begin{eqnarray}
  \Num_{\rm C}(\EWr) & = & \sum_{W_{\rm i} \ge \EWr - 0.5\Delta \EWr}^{W_{\rm i} < \EWr +
    0.5\Delta \EWr} \frac{1} {C(W_{\rm i})} \label{eqn.numc} \\
\sigma_{\Num_{\rm C}}^2 & = & \sigma_{\Num_{\rm obs}}^{2} +
\sum_{W_{\rm i} \ge \EWr - 0.5\Delta \EWr}^{W_{\rm i} < \EWr + 
    0.5\Delta \EWr} \bigg( \frac{ \sigma_{C(W_{\rm i})} } { C(W_{\rm
      i})^2 } \bigg)^2 {\rm .} \nonumber 
\end{eqnarray}
Again, $\Num_{\rm obs}$ is the contribution of the actual observed
number of absorbers in the given \EWr\ bin, and the
completeness-corrected number $\Num_{\rm C} \ge \Num_{\rm
  obs}$. The error $\sigma_{\Num_{\rm obs}}$ is estimated from a
Poisson distribution if $\Num_{\rm obs} < 120$ and from a
Gaussian approximation if $\Num_{\rm obs} \ge 120$.

We subtract \dNFkIVdX\ from all quoted \dNCIVdX\ values, in the
appropriate redshift bins (see \S\ref{subsec.biases}) and add the
errors in quadrature. For $\EWr \ge 0.6\Ang$, $\dNFkIVdX =
0.020^{+0.003}_{-0.002}$ ($1.4 < z < 4.6$); $0.022^{+0.005}_{-0.004}$
($z < 2.97$); and $0.020^{+0.004}_{-0.003}$ ($z \ge 2.97$).

We present \dNCIVdX\ for different equivalent width limits in Figure
\ref{fig.dndx}. The line density shows little differential evolution
based on \EWlin{\rm lim}, as expected from the consistent shape of the
frequency distributions over time (Figure \ref{fig.fw}). For $\EWr
\ge 0.6\Ang$, the inverse-variance weighted average $\bar{\dNCIVdX} =
0.350\pm0.005$ for $1.46 \le z < 1.96$ (or the lowest five redshift
bins). This average is $2.37\pm0.09$ times larger than the highest
redshift \dNCIVdX\ at $z = 3.76$.  Though the magnitude of the
increase is modest, the detection is a $>20\sigma$ result. Thus, the
line density grows consistently and smoothly from $\zciv = 4.5
\rightarrow\ \approx 1.74$, then plateaus at $\dNCIVdX \approx 0.34$
until $z = 1.46$, as expected from the best-fit \ff{\EWr}\ parameters
(see Figure \ref{fig.fwellipse}).

There is a known bias in the SDSS quasar color selection that leads to
an excess of Lyman-limit systems at $3 \lesssim \zqso \lesssim 3.5$
\citep{prochaskaetal10, worseckandprochaska11}, which likely increases
the incidence of strong metal-line absorption systems at these
redshifts. Therefore, there is potential for the decrease in the
highest SDSS bin to be even larger, but the effect of the color
selection on the \ion{C}{4} sample is beyond the scope of this paper.

\begin{figure*}[hbt]
  \begin{center}$
    \begin{array}{cc}
  \includegraphics[width=0.47\textwidth]{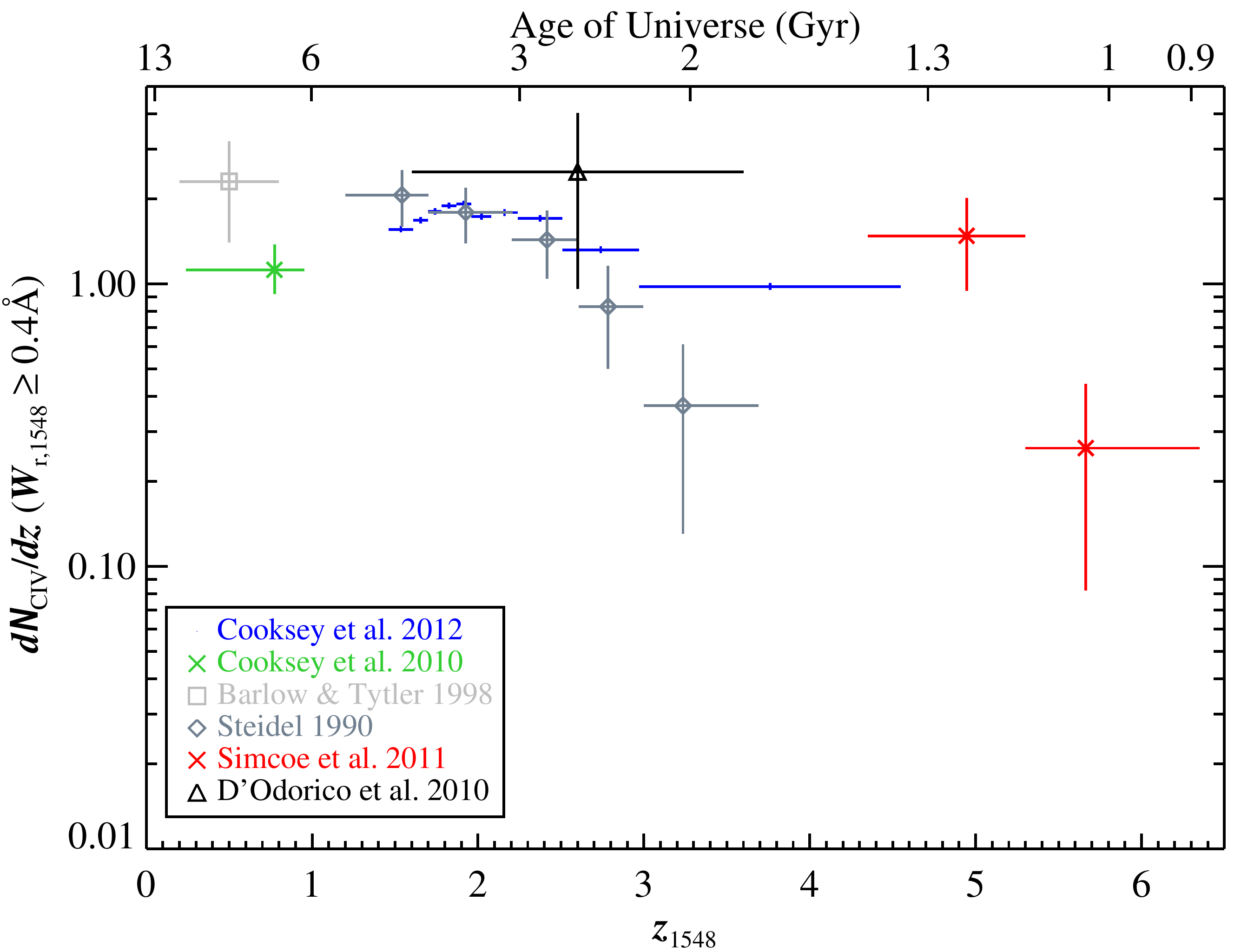}  & 
  \includegraphics[width=0.47\textwidth]{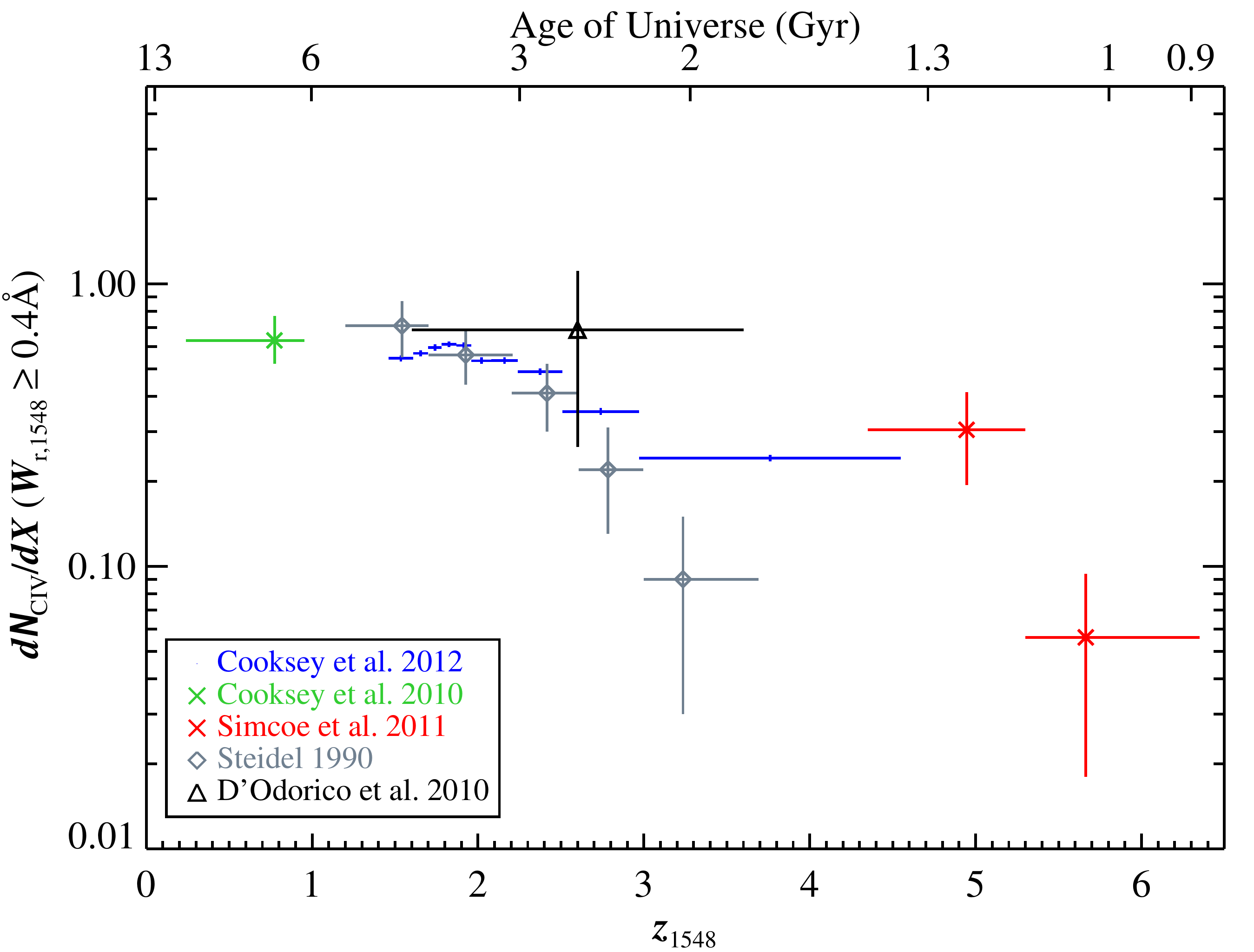} 
  \end{array}
 $\end{center}
  \caption[Redshift and co-moving \ion{C}{4} absorber line densities.]
  {Redshift (left) and co-moving (right) \ion{C}{4} absorber line
    densities. We compare apples-to-apples \dNCIVdz\ with $\EWr \ge
    0.4\Ang$ with five surveys: \citet[][light gray
    square]{barlowandtytler98}, \citet[][green cross]{cookseyetal10},
    \citet[][black triangle]{dodoricoetal10}, \citet[][gray
    diamonds]{steidel90}, and \citet[red crosses]{simcoeetal11}. The
    co-moving line density steadily increases from high to low
    redshift.  The growth in \dNCIVdX\ indicates a consistent increase
    with time in the product of the co-moving number density and the
    physical cross-section of the absorbing clouds.
    \label{fig.dndzlit}
  }
\end{figure*}

\subsection{Comparison with Previous Results}\label{subsec.prevrslt}

To fairly compare to other surveys, we applied various \EWr\ cuts to
our complete sample, chosen to match the corresponding cuts of prior
surveys. 

To compare \ff{\EWr}, we converted published column density frequency
distributions {\it fits} from \citet{songaila01} and
\citet{dodoricoetal10} to equivalent width frequency distributions by
assuming the linear curve of growth and mapping \ff{\NCIV}\ directly
to \ff{\EWlin{1548}}: $\ff{\EWr} = \ff{\NCIV} \ud \NCIV / \ud
\EWr$. We adjusted for cosmology as necessary.  A single-component
cloud becomes saturated and nonlinear at $\logCIV \approx 14$, which
translates to $\EWlin{1548} \approx 0.6\Ang$, where we typically are
$\approx 50\%$ complete.  

While we have good statistics on the rare, strong absorbers, we suffer
from incompleteness at $\EWr \lesssim 0.6\Ang$. \citet{songaila01} and
\citet{dodoricoetal10} were smaller, higher-resolution, higher-S/N
studies, and they were complete to very low \EWr\ but suffered from
sample variance at larger equivalent widths. However, Figure
\ref{fig.fwlit} shows that our results are consistent with each other
in the overlap region. Our best-fit parameters to the exponential
\ff{\EWr} in the \citet{dodoricoetal10} and \citet{songaila01}
redshift bins were, respectively:
$\alpha=-2.65^{+0.04}_{-0.04}\Ang^{-1}$ and
$k=3.72^{+0.37}_{-0.35}\Ang^{-1}$ for $1.6 \le z \le 3.6$ with 7884
systems; and $\alpha=-2.58^{+0.13}_{-0.13}\Ang^{-1}$ and
$k=2.29^{+0.80}_{-0.69}\Ang^{-1}$ for $2.9 \le z \le 3.54$ with 878
systems.

We compared the SDSS redshift and co-moving line densities with the
literature \citep{steidel90, barlowandtytler98, cookseyetal10,
  dodoricoetal10, simcoeetal11}. For \citet{dodoricoetal10}, we used
their best-fit \ff{\NCIV}\ values to calculate \dNCIVdX\ and $\dNCIVdz
= (\dNCIVdX)(\dXdz)$ for $\logCIV \ge 14$. For
\citet{barlowandtytler98} and \citet{steidel90}, we estimated
$\dNCIVdX\ = (\dNCIVdz)(\dXdz)^{-1}$. The \citet{steidel90} results
match ours well for the range where the author was fairly complete but
with $\approx20\%$ uncertainty, compared to our $\approx2\%$ errors
(see Figure \ref{fig.dndzlit}). Extending the redshift
coverage by including the low-redshift measurements of
\citet{barlowandtytler98} and \citet{cookseyetal10} and the
high-redshift values from \citet{simcoeetal11}, we see that \dNCIVdX\
has steadily increased by, roughly, a factor of ten from $\zciv = 6
\rightarrow 0$.

\section{Discussion}\label{sec.discuss}

\subsection{\ion{C}{4} Evolution\label{subsec.civevo}}

\begin{figure}[hbt]
  \begin{center}$
    \begin{array}{ll}
      \hspace{1.5mm}\includegraphics[width=0.43\textwidth]{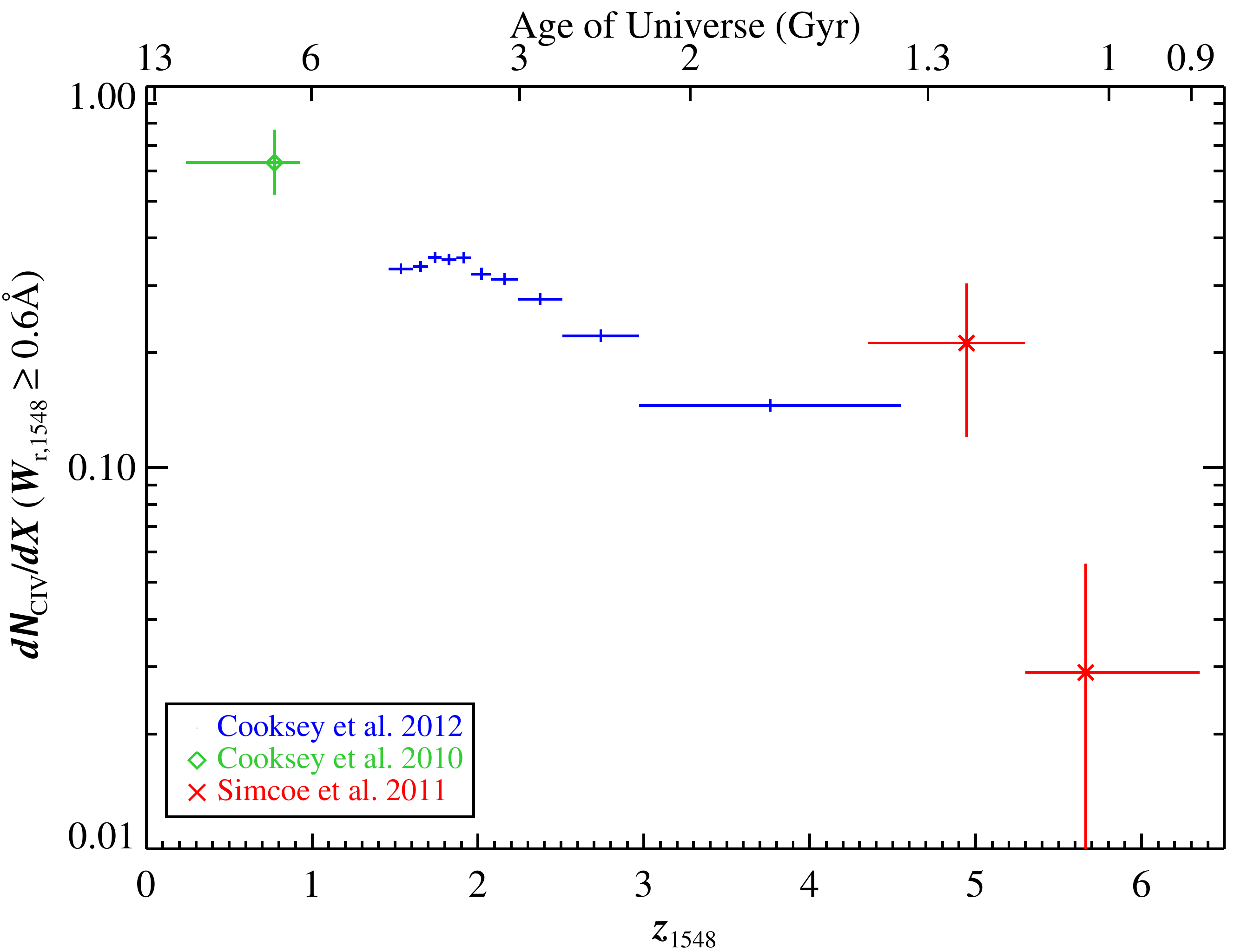}  \\
      \includegraphics[width=0.47\textwidth]{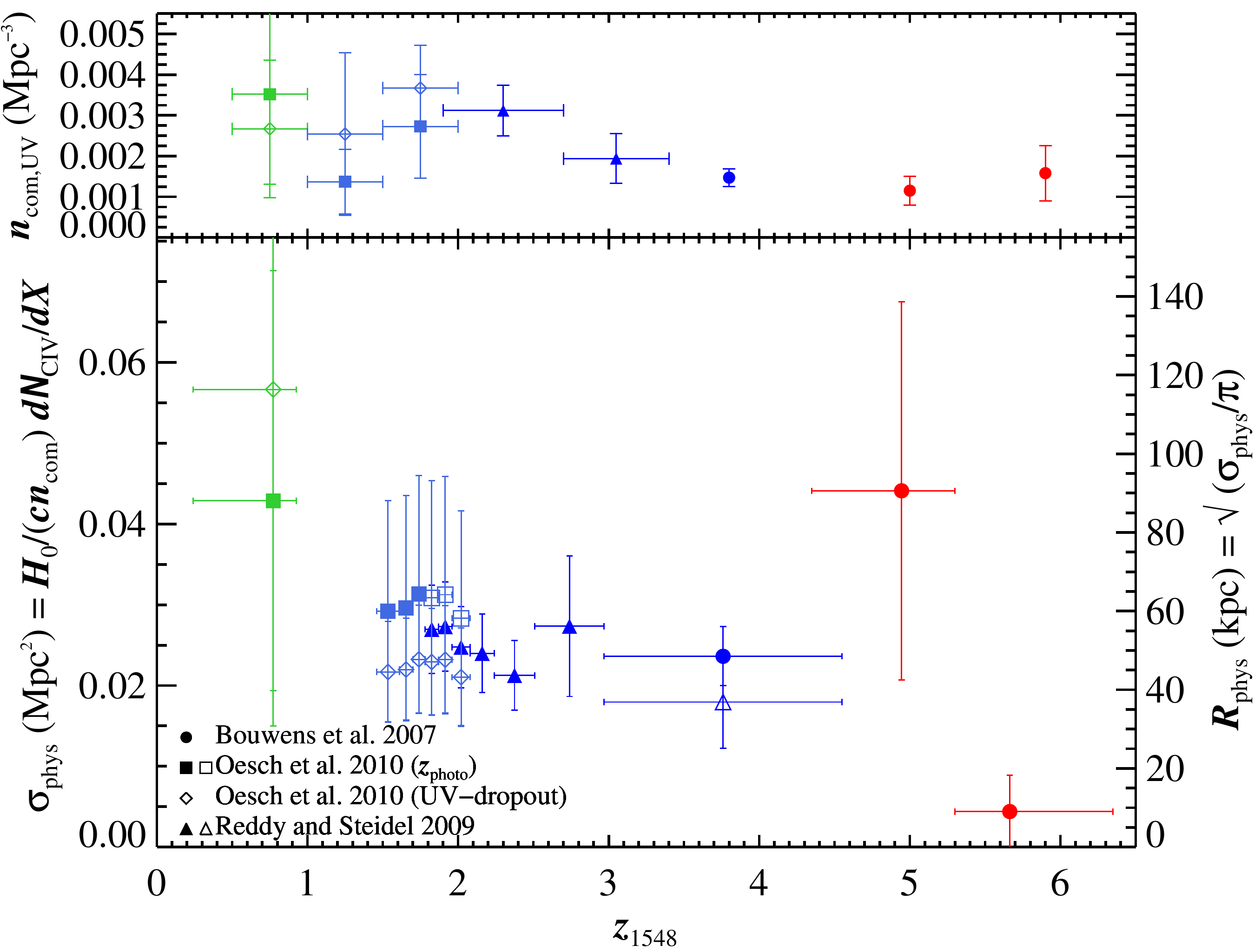}  
    \end{array}$
  \end{center}
  \caption[Evolving nature of \ion{C}{4} absorbers.]
  {Evolving nature of \ion{C}{4} absorbers. {\it Top}: By combining
    the current SDSS values with those from \citet{cookseyetal10} and
    \citet{simcoeetal11}, we show that \dNCIVdX\ for $\EWlin{1548} \ge
    0.6\Ang$ has increased 10-fold, roughly, from $z = 6 \rightarrow
    0$. {\it Middle}: We calculated the co-moving number density of
    UV-selected galaxies by integrating the UV luminosity functions
    from \citet{bouwensetal07}, \citet{oeschetal10}, and
    \citet{reddyandsteidel09}, down to $0.5\,L^{\ast}$. UV-selected
    galaxies are bright, star-forming galaxies, such as the LBGs that
    \citet{adelbergeretal05} showed likely host most $\logCIV \ge 14$
    absorbers at $2 \lesssim z \lesssim 3$. {\it Bottom}: Assuming
    that all $\EWlin{1548} \ge 0.6\Ang$ doublets are in UV-selected
    galaxy halos, we estimated the galaxy-\ion{C}{4} cross-section as
    a function of redshift with Equation \ref{eqn.dndxphys}. The
    inferred cross-sections from the SDSS $1.5 \lesssim z \lesssim
    4.5$ sample are consistent with the area that
    \citet{adelbergeretal05} and \citet{steideletal10} measured.
    \label{fig.dndxbest}
  }
\end{figure}

We show the best measurements of \dNCIVdX\ for $\EWr \ge 0.6\Ang$ and
$0 < z < 6.5$ in the top panel of Figure
\ref{fig.dndxbest}.\footnote{For $z < 1$, we summed the eight $\EWr
  \ge 0.6\Ang$ systems in the {\it HST} sample \citep{cookseyetal10}
  to get $\dNCIVdX = 0.63^{+0.14}_{-0.11}$.  For $z > 4$, we have
  $\dNCIVdX = 0.212\pm0.092$ for $z = 4.94$ (2 systems) and
  $0.026\pm0.027$ for $z = 5.66$ (1) from the FIRE sample
  \citep{simcoeetal11}.}  The co-moving line density relates to the
co-moving volume density of absorbing clouds \ncom\ and their physical
cross-section \sigphys:
\begin{equation}
\frac{\displaystyle \ud \Num_{\Cthr}}{\displaystyle \ud X} =
\frac{\displaystyle c}{\displaystyle H_{0}} \ncom \sigphys {\rm .} \label{eqn.dndxphys} 
\end{equation}
Thus, the roughly order-of-magnitude increase in \dNCIVdX\ from
high-to-low redshift means the product \ncom\sigphys\ has increased by
a factor of $\approx10$. We know the metallicity of the universe has
steadily increased over cosmic time, so the possible number of
\ion{C}{4}-absorbing clouds (i.e., \ncom) has likely increased. At
least some \ion{C}{4} absorption traces galaxy halos at low
\citep{chenetal01} and high redshift \citep{adelbergeretal05,
  martinetal10, steideletal10}. Since galaxies (and their halos) have
likely grown over cosmic time, increases in both \sigphys\ and \ncom\
appear to contribute to the increase in \dNCIVdX.

\citet{adelbergeretal05} reported that almost all $\logCIV \ge 14$
absorbers arise within $\approx80\kpc$ of Lyman-break galaxies (LBGs)
at $2 \lesssim z \lesssim 3$. The evidence included: individual strong
absorber-LBG pairs; strong absorption in stacked spectra of close
background galaxies, shifted to the rest-frame of the foreground
galaxies; and similar LBG-\ion{C}{4} cross-correlation and LBG
autocorrelation functions, suggesting they have the same spatial
distribution. 

\citet{steideletal10} increased the LBG-LBG pairs for stacking
analysis, and they measured average $\EWlin{1548} = 0.13\pm0.05\Ang$
and $1.18\pm0.15\Ang$ at distances $b = 63\kpc$ and 103\kpc,
respectively. Interpolating between these two measurements, an average
$\EWlin{1548} = 0.6\Ang$ system would reside at $b \approx
85\kpc$. The \citet{steideletal10} LBG sample went as faint as
$\approx 0.3\,L^{\ast}$, assuming the luminosity function of
\citet{reddyandsteidel09}, and since they used stacks of galaxy
spectra, their analysis included the effects of partial covering
fractions.

Since the $\EWr \ge 0.6\Ang$ \ion{C}{4} absorbers in our sample have
$\logCIV \gtrsim 14$, we used the co-moving number density of
UV-selected galaxies, to
estimate the typical galaxy-\ion{C}{4} cross-section over time with
Equation \ref{eqn.dndxphys}.  We measured \ncomuv\ with the UV
luminosity functions from \citet{oeschetal10},
\citet{reddyandsteidel09} and \citet{bouwensetal07}, which covered
several smaller redshift bins spanning $0.5 < z < 2$, $1.9 < z < 3.4$,
and $3.8 \lesssim z \lesssim 5.9$, respectively. For each luminosity
function, we integrated the best-fit Schechter function down to
$0.5\,L^{\ast}$ or $\approx0.75\,$mag fainter than the published
$M_{\rm UV}^{\ast}$, which ranged between $\approx-19$ and
$-21\,$mag. We estimated the \ncomuv\ errors with Monte Carlo
simulations.

The UV-selected galaxy number density, \ncomuv, may increase by a factor of two
to three from $z \approx 6 \rightarrow 0$, but the uncertainties are
large (see Figure \ref{fig.dndxbest}, middle panel). Applying
\ncomuv\ and \dNCIVdX\ to Equation \ref{eqn.dndxphys}, we estimated
\sigphys, the physical galaxy-\ion{C}{4} cross-section (Figure
\ref{fig.dndxbest}, lower panel). For some redshift bins, multiple
luminosity functions could be used in conjunction with our \dNCIVdX,
and we show all resulting values but highlight the preferred values
with filled symbols.

The galaxy-\ion{C}{4} cross-section shows no evolution over the SDSS
redshift range within the errors, which are dominated by the
20\%--60\% uncertainties in \ncomuv. Assuming the
cross-section is due to a spherical halo that projects with 100\%
\ion{C}{4} covering fraction, the halo radius would be $R_{\rm phys} =
\sqrt{\sigphys/\pi} \approx 50\kpc$ for $1.5 \lesssim z
\lesssim 4.5$. This distance agrees with \citet{adelbergeretal05} and
allows for e.g., a non-unity covering fraction or limiting to brighter
UV-selected galaxies, both of which would increase $R_{\rm phys}$.

There is an approximately 10-fold increase in \sigphys\ from $z
\approx 6 \rightarrow 0$ when we include the \dNCIVdX\ measurements
from \citet{simcoeetal11} and \citet{cookseyetal10},
respectively. This increase in the galaxy-\ion{C}{4} cross-section is
comparable to that of \dNCIVdX\ over the same redshift range.  Thus,
if the $\EWr \ge 0.6\Ang$ \ion{C}{4} absorbers were only tracing galaxy
halos, the redshift evolution of \dNCIVdX\ could be solely due to the
halos filling up with triply-ionized carbon, through some combination
of physical growth, increased metallicity, and\slash or evolution in
the ionizing background.

However, the uncertainty in the $z < 1$ cross-section estimate is
large and consistent with an increase of only two to three compared
with the $z \approx 6$ value. In this case, the roughly
order-of-magnitude increase in \dNCIVdX\ could equally be due to
modest increases in \ncomuv\ and \sigphys. 

We emphasize that the preceding discussion assumes all $\EWr \ge 0.6$
\ion{C}{4} systems at $0 < z < 6$ are only found in the
circum-galactic media of UV-selected galaxies. Any intergalactic
\ion{C}{4} contribution to \dNCIVdX\ at any redshift would cause us to
overestimate the galaxy-\ion{C}{4} cross-section.

Evolution in the ultraviolet background (UVB) possibly explains the
seeming spike in \sigphys\ at $z \approx 5$. \citet{simcoe11} showed
that \ion{C}{4} is a preferred, if not dominant, transition of carbon
at $z = 4.3$ compared to lower redshift, assuming the UVB of
\citet{haardtandmadau01} or \citet{fauchergiguereetal09}. Though the
redshift window of this effect is small ($\Delta z \approx 0.5$), the
increase in \ion{C}{4}-ionizing photons would increase the \ion{C}{4}
cross-section of UV-selected galaxies and may explain the spike in
\dNCIVdX\ and \sigphys. In general, triply-ionized carbon increasingly
becomes a disfavored transition of carbon with decreasing redshift,
given a standard model of the UVB \citep{oppenheimeranddave08,
  simcoe11}. The UVB likely dominates the ionization flux at the
expected tens of kiloparsecs from the host galaxies. Thus, the
increasing number of \ion{C}{4} absorbers towards lower redshift
indicates an increasing enrichment of the gas, to more-than-compensate
for the decreasingly favored triply-ionized state.

\ion{C}{4} absorption could also be a galactic wind signature
\citep[see][]{steideletal10}. However, the \dNCIVdX\ redshift
evolution does not mimic the evolution of \dNMgIIdX\ for $\EWlin{2796}
\gtrsim 1\Ang$ absorbers, and these strong \ion{Mg}{2} doublets are
often analyzed as wind tracers \citep[][but also see
\citealt{gauthieretal10} and \citealt{kacprzaketal11}]{weineretal09,
  rubinetal10, bordoloietal11}. Strong \ion{Mg}{2} absorbers evolve
strongly with redshift by increasing with increasing redshift
\citep{nestoretal05, prochteretal06}, peaking at $z \approx 3$, and
then decreasing at higher redshifts \citep{matejekandsimcoe12ph}. The
latter study showed that the evolution in \dNMgIIdX\ tracks the cosmic
star-formation rate \citep{bouwensetal10}, using the scaling relation
of \citet{menardetal11}, reinforcing the idea that strong \ion{Mg}{2}
absorbers arise in galactic winds.

We see no decrease at lower redshifts for strong \ion{C}{4} absorbers
(see Figure \ref{fig.dndx}). If strong \ion{C}{4} and \ion{Mg}{2}
absorbers were both tracing winds, then, to account for their
different redshift evolution, either: they must probe outflows in
different ways; or strong \ion{C}{4} absorption traces an additional
medium, such as halo gas, that contributes significantly to its
\dNCIVdX\ evolution.


\subsection{\ion{C}{4} Mass Density}\label{subsec.omciv}

The \ion{C}{4} doublet redshifts into optical passbands for $1.5
\lesssim z \lesssim 5.5$ and has been observed extensively with large,
ground-based telescopes \citep{songaila01, boksenbergetal03ph,
  schayeetal03, scannapiecoetal06, dodoricoetal10}.  Surveys of
ultraviolet quasar spectra taken with {\it HST} cover the $z < 1$
\ion{C}{4} systems \citep{danforthandshull08, cookseyetal10}. More
recently, improved infrared spectrographs have pushed \ion{C}{4}
surveys to $z \approx 6$ \citep{ryanweberetal09, beckeretal09,
  simcoeetal11}. Typically these studies have focused on the evolution
of \OmCIV, the \ion{C}{4} mass density relative to the closure
density.  The first moment of the column density frequency
distribution \ff{\NCIV} is related to \OmCIV\ as follows:
\begin{equation} \OmCIV =
  \frac{H_{0}\,m_{\rm C}} {c\,\rho_{c,0}}
  \int_{\N{\rm min}}^{\N{\rm max}} \ff{\NCIV} \NCIV \ud \NCIV {\rm
    ,} \label{eqn.omciv}
\end{equation}
where $H_{0} = 71.9\kms\,\mathrm{Mpc}^{-1}$ is the Hubble constant
today; $m_{\rm C} =2\times10^{-23}\,$g is the mass of the carbon atom;
$c$ is the speed of light; and the critical density
$\rho_{c,0}=3H_{0}^2(8\pi G)^{-1} =9.77\times10^{-30}\,{\rm g}\cm{-3}$
for our assumed Hubble constant.

The earliest (optical) studies found that \OmCIV\ was relatively
constant for $2 \lesssim z \lesssim 4.5$ \citep{songaila01,
  boksenbergetal03ph, schayeetal03, songaila05,
  scannapiecoetal06}. \citet{cookseyetal10} showed that \OmCIV\
increased by at least a factor of four from $z \approx 3$ to $z <
1$. Using better optical spectra, \citet{dodoricoetal10} found that
\OmCIV\ increased smoothly from $z = 3$ to $z = 1.5$ and mapped well
onto the $z < 1$ values.

However, all these studies measured a power-law shape for the
\ff{\NCIV}, which formally corresponds to an infinite \ion{C}{4} mass
density. These surveys were limited to small numbers of available
sightlines, typically less than 50. The rarest systems are the
strongest, and these dominate the mass density measurement when the
distribution is a power law. Hence the small-number statistics on the
strongest \ion{C}{4} absorbers limited the quoted \OmCIV\ to be for
$12 \lesssim \logCIV \lesssim 15$, with one survey pushing to $\logCIV
\approx 16$ \citep{scannapiecoetal06}. Indeed, scaling the
\citet{scannapiecoetal06} \OmCIV\ to $\logCIV \le 15$ reduces their
value by 45\% \citep{cookseyetal10}.

Our analysis, however, provides good statistics on the rare, strong
systems, which dominate the SDSS sample.  The observed \OmCIV\ can be
approximated by the sum of the detected \ion{C}{4} absorbers
\citep{lanzettaetal91}. The total column density in a given redshift
bin is simply:
\begin{eqnarray}
\N{\rm tot} & = & \sum_{\Num_{\rm obs}} \NCIV \label{eqn.ncolmtot} \\
\sigma_{\N{\rm tot}}^{2} & = & \sigma_{\Num_{\rm obs}}^{2} +
\sum_{\Num_{\rm obs}} \sigNCIV^{2} {\rm ,} \nonumber 
\end{eqnarray}
where, again, we factor in the counting uncertainty for the number of
detections $\sigma_{\Num_{\rm obs}}$. Then, we estimated the mass
density relative to the critical density as:
\begin{eqnarray}
  \OmCIV & = & \frac{H_{0}\,m_{\rm C}} {c\,\rho_{c,0}}
  \frac{\displaystyle \N{\rm tot}}
  {\displaystyle \langle \DX{\EWr} \rangle}  \label{eqn.omciv_sum}
  \\ 
  \sigOmCIV^{2} & = & \OmCIV^{2} \Bigg( \bigg( \frac{\displaystyle
    \sigma_{\N{\rm tot} }}{ \displaystyle \N{\rm tot} } \bigg)^{2} + 
\bigg( \frac{\displaystyle \sigma_{\langle \DXp \rangle} }{
\displaystyle \langle \DX{\EWr} \rangle } \bigg)^{2} \Bigg)  {\rm .}
\nonumber 
\end{eqnarray}
Since the completeness corrections were compiled in equivalent width
space, we used the median \DX{\EWr} available to each of the
absorbers in the given bin, and the error was the standard deviation
of the detections. The small scaling to account for the accepted
false-positive distribution follows that of the frequency distribution
(Equation \ref{eqn.afpfw}).

\begin{figure}[hbt]
  \begin{center}
  \includegraphics[width=0.47\textwidth]{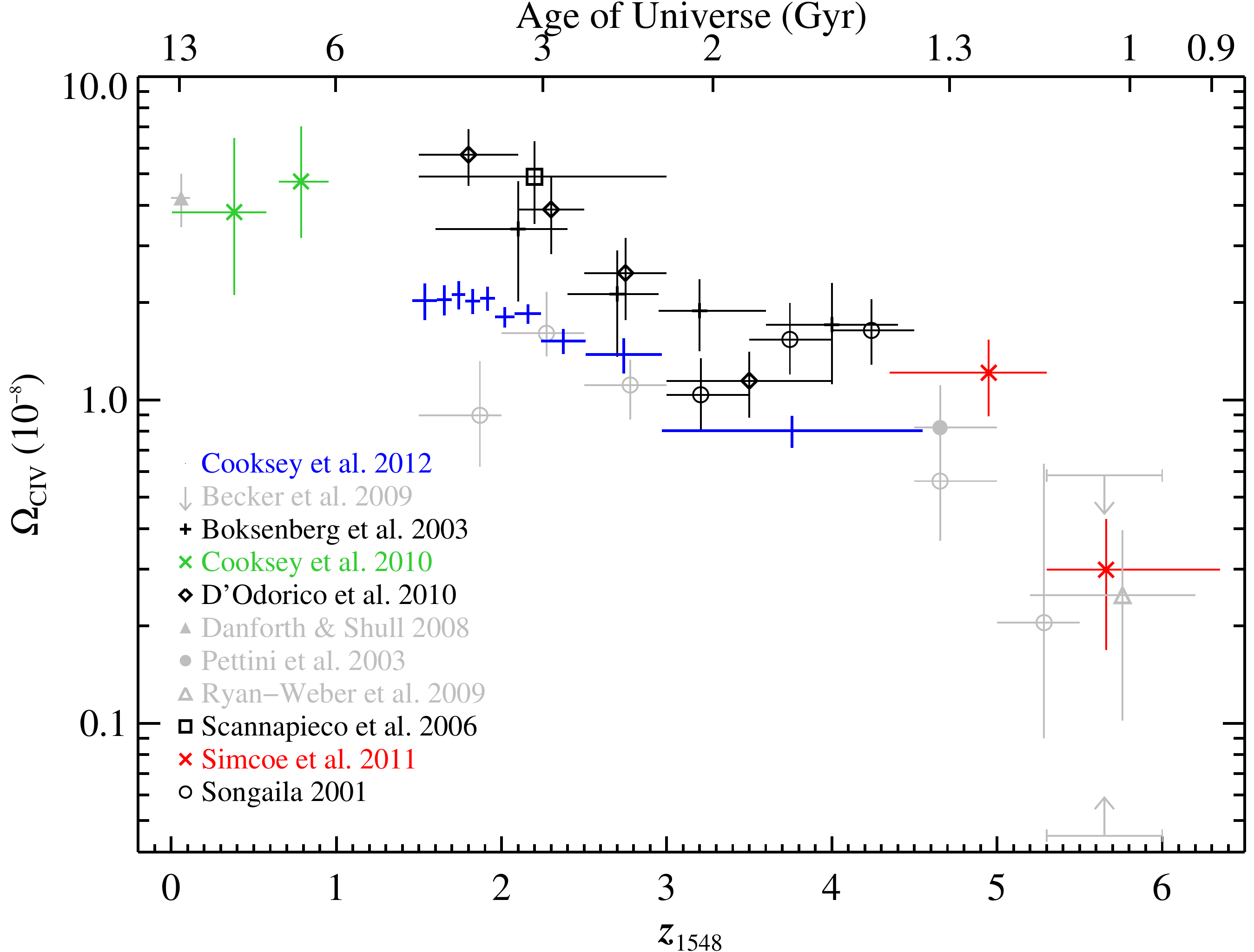} 
  \end{center}
  \caption[Redshift evolution of \OmCIV\ for $\logCIV \ge 14$ absorbers.]
  {Redshift evolution of \OmCIV\ for $\logCIV \ge 14$ absorbers. The
    SDSS apparent optical depth values are {\it lower} limits (blue),
    since most systems are saturated. However, these lower limits
   firmly exclude the $z < 3$ measurements from \citet{songaila01},
    which have been grayed out with the other superseded measurements,
    and support the result that \OmCIV\ smoothly increases from $z
    \approx 3$ to $z = 1.5$ from \citet{dodoricoetal10}.
    \label{fig.omciv}
  }
\end{figure}

In Figure \ref{fig.omciv}, we plot \OmCIV\ over redshift from several
studies for $\logCIV \ge 14$, adjusted for the new limits and
cosmology \citep[see ][for details]{cookseyetal10}. The direct SDSS
measurements are {\it lower} limits since we have predominately
saturated absorbers and use the AODM to estimate column
densities. However, we can still exclude values below the limits,
including the \citet{songaila01} measurements at $z <
3$.\footnote{\citet{schayeetal03} and \citet{songaila05} used pixel
  optical depth methods to also measure \OmCIV, but it is unclear how best
  to compare their values to our traditional quasar absorption-line
  study. For this reason, we omit discussion of their results.}
This reinforces the \citet{dodoricoetal10} result that
\OmCIV\ smoothly increases.  In an upcoming paper, we will be
combining the \citet{dodoricoetal10} and current datasets to fit
\ff{\NCIV}\ and measure \OmCIV\ for $\logCIV \gtrsim 12$ via Equation
\ref{eqn.omciv}.

\section{Summary}\label{sec.summ}

We have completed the largest \ion{C}{4} survey to date by leveraging
the SDSS DR7 quasar catalog \citep{schneideretal10}. In 26,030
sightlines with $\zqso \ge 1.7$ and $\langle {\rm S/N} \rangle \ge
4\,{\rm pixel}^{-1}$, we identified 14,772 \ion{C}{4} systems at least
5000\kms\ blueward of the quasar. The entire doublet catalog and other
tools for analysis (e.g., completeness grids) are made available for the
community.\footnote{See
  \url{http://igmabsorbers.info/}.}

The bulk of the \ion{C}{4} survey was automated, from continuum
fitting to candidate selection. We visually verified all doublets in
the final catalog. Our Monte Carlo completeness tests included the
effects of the automated algorithms, user bias, and accepted false
positives. We analyzed the sample as a whole and in ten redshift bins
of roughly 1500 doublets each. We also constructed and analyzed other
subsamples as needed, for specific comparisons to published studies.

The equivalent width frequency distribution \ff{\EWr}\ is described
well by an exponential model. The best-fit parameters evolved smoothly
with redshift, with the largest change arising in the
normalization. The parameter evolution follows the trend of increasing
\dNCIVdX\ with increasing redshift---a $2.37\pm0.09$ increase over $z
= 4.55 \rightarrow\ \approx\!1.74$. For $1.46 \le z < 1.96$, $\dNCIVdX
\approx 0.34$.  

In the region of overlap, the SDSS \ff{\EWr}\ agree well with the
published column density frequency distributions \ff{\NCIV}\ from
smaller, high-resolution, high-S/N surveys \citep{songaila01,
  dodoricoetal10}. We converted \ff{\NCIV} to \ff{\EWr}\ assuming the
linear curve-of-growth. The published \ff{\NCIV}\ distributions were
best fit by a power-law formalism. Thus, combined with the exponential
nature of the strong-end of \ff{\EWr}, we at last see an exponential
cut-off in the equivalent width distribution that has not been
previously characterized because of small sample sizes. The location
of this cut-off sets the high-end range for \ion{C}{4} absorbers and
formally leads to convergence in the cosmic \ion{C}{4} mass density.

In this paper, we joined our large sample of strong absorbers to the
fits of weak absorbers obtained at high resolution. In a forthcoming
paper, we will perform a joint analysis of our SDSS sample with the
high-resolution data of \citet{dodoricoetal10} to fit clean forms to
the full range of \ff{\NCIV}. This will allow us to improve our
understanding of the \ion{C}{4} mass density evolution, where
currently our SDSS study only yields lower limits for $\logCIV \ge 14$
systems.

The co-moving line density \dNCIVdX\ smoothly increases,
approximately, ten-fold from $z = 6 \rightarrow 0$ for $\EWlin{1548}
\ge 0.6\Ang$ absorbers. Physically, this means that the physical cross
section (\sigphys) and\slash or co-moving number density of \ion{C}{4}
absorbing clouds (\ncom) has increased steadily over time, since
\dNCIVdX\ is proportional to their product. We estimated the projected
\ion{C}{4}-absorbing cross-section of UV-selected galaxies to be
$R_{\rm phys} \approx 50\,$kpc, by assuming all the SDSS $\EWr \ge
0.6\Ang$ absorbers arise in $\ge 0.5\,L^{\ast}$ galaxy halos, which is in
good agreement with \citet{adelbergeretal05} and
\citet{steideletal10}.

This is the first paper in a series on various metal lines in the
SDSS DR7 quasar spectra, which will provide a comprehensive and
uniform collection of catalogs for future analysis and detailed
comparison with cosmological hydrodynamic simulations.


\acknowledgements We thank J. Harker, D. Ogden, J. Pritchard, and
P. Creasey for their programming help and P. Jonsson for productive
discussions regarding statistics and programming. We gratefully
acknowledge the vital role of the Adam J. Burgasser Endowed Chair.
The current study was funded largely by the National Science
Foundation Astronomy \& Astrophysics Postdoctoral Fellowship
(AST-1003139) and in part by the MIT Department of Physics and the
Alfred P. Sloan Foundation Research Fellowship.

Funding for the SDSS and SDSS-II has been provided by the Alfred
P. Sloan Foundation, the Participating Institutions, the National
Science Foundation, the U.S. Department of Energy, the National
Aeronautics and Space Administration, the Japanese Monbukagakusho, the
Max Planck Society, and the Higher Education Funding Council for
England. The SDSS Web Site is \url{http://www.sdss.org/}.

The SDSS is managed by the Astrophysical Research Consortium for the
Participating Institutions. The Participating Institutions are the
American Museum of Natural History, Astrophysical Institute Potsdam,
University of Basel, University of Cambridge, Case Western Reserve
University, University of Chicago, Drexel University, Fermilab, the
Institute for Advanced Study, the Japan Participation Group, Johns
Hopkins University, the Joint Institute for Nuclear Astrophysics, the
Kavli Institute for Particle Astrophysics and Cosmology, the Korean
Scientist Group, the Chinese Academy of Sciences (LAMOST), Los Alamos
National Laboratory, the Max-Planck-Institute for Astronomy (MPIA),
the Max-Planck-Institute for Astrophysics (MPA), New Mexico State
University, Ohio State University, University of Pittsburgh,
University of Portsmouth, Princeton University, the United States
Naval Observatory, and the University of Washington.

{\it Facilities:} \facility{Sloan}

\bibliographystyle{apj} \bibliography{../../sdss}

\end{document}